\documentclass[11pt,a4paper]{article}
\pdfoutput=1
\usepackage{jheppub} 
\usepackage{xspace}
\usepackage{subfigure}
\usepackage{multirow}

\newcommand{\ra} {\rightarrow}

\newcommand{\be}{\begin{equation}}
\newcommand{\ee}{\end{equation}}

\newcommand{\bea}{\begin{eqnarray}}
\newcommand{\eea}{\end{eqnarray}}
\newcommand{\beanon}{\begin{eqnarray*}}
\newcommand{\eeanon}{\end{eqnarray*}}
\newcommand{\ba}{\begin{array}}
\newcommand{\ea}{\end{array}}
\newcommand{\bd}{\begin{description}}
\newcommand{\ed}{\end{description}}
\newcommand{\bi}{\begin{itemize}}
\newcommand{\ei}{\end{itemize}}
\newcommand{\ben}{\begin{enumerate}}
\newcommand{\een}{\end{enumerate}}
\newcommand{\bc}{\begin{center}}
\newcommand{\ec}{\end{center}}

\newcommand{\GeV}{\mbox{${\mathrm GeV}$}\xspace}

\newcommand{\tw}{\theta_{\mathrm{w}}}

\newcommand{\ordEW}{\mathcal{O}(\alpha_{\scriptscriptstyle EM}^6)\xspace}
\newcommand{\ordQCD}{\mathcal{O}(\alpha_{\scriptscriptstyle EM}^4
  \alpha_{\scriptscriptstyle S}^2)\xspace}
\newcommand{\ordQCDsq}{\mathcal{O}(\alpha_{\scriptscriptstyle EM}^2
  \alpha_{\scriptscriptstyle S}^4)\xspace}

\newcommand{\eqn}[1]{Eq.(\ref{#1})}
\newcommand{\eqns}[2]{Eqs.(\ref{#1}--\ref{#2})}

\newcommand{\tbn}[1]{Tab.~\ref{#1}}

\newcommand{\fig}[1]{Fig.~\ref{#1}}

\newcommand{\sect}[1]{Sect.~\ref{#1}}

\newcommand{\subsect}[1]{Sub--Sect.~\ref{#1}}

\newcommand{\rf}[1]{Ref.~\cite{#1}}
\newcommand{\rfs}[1]{Refs.~\cite{#1}}  

\newcommand{\Phantom}{{\tt PHANTOM}\xspace}

\newcommand{\MadEvent}{{\tt MADEVENT}\xspace}

\newcommand{\Whizard}{{\tt WHIZARD}\xspace}

\newcommand{\mc}[1]{\mathcal{#1}}
\newcommand{\sla}[1]{\displaystyle{\not} #1}

\newcommand{\ifb}{\mbox{ fb}^{-1}}

\newcommand{\gb}{\omega}

\newcommand{\de}{\text{d}}
\newcommand{\Imm}{\text{Im}\,}
\newcommand{\Ree}{\text{Re}\,}

\title{
Vector Boson scattering at the LHC: counting experiments for unitarized models 
in a full six fermion approach.
}
 
\author[a]{Alessandro Ballestrero,}
\author[a,b]{Diogo Buarque Franzosi,}
\author[b,c]{Luisa Oggero}
\author[a,b]{and Ezio Maina}

\affiliation[a]{INFN, Sezione di Torino,\\
Via Giuria 1, 10125 Torino, Italy}
\affiliation[b]{Dipartimento di Fisica Teorica, Universit\`a di Torino,\\
Via Giuria 1, 10125 Torino, Italy}
\affiliation[c]{Albert-Ludwigs-Universit\"{a}t Freiburg, Physikalisches Institut,\\
D-79104 Freiburg, Germany}

\emailAdd{ballestrero@to.infn.it}
\emailAdd{buarque@to.infn.it}
\emailAdd{maina@to.infn.it}
\emailAdd{luisa.oggero@physik.uni-freiburg.de}

\preprint{DFTT 31/2011, FR-PHENO-2011-023}

\arxivnumber{1112.1171}

\abstract{
Unitarization models describe phenomenologically the high energy behaviour
of a strongly interacting symmetry breaking sector.
In this work, predictions of some unitarized models in vector boson scattering at LHC are 
studied and compared with analogous studies in Equivalent Vector Boson 
Approximation and previous results for  the benchmark no-Higgs scenario. 
To perform such studies, unitarized model amplitudes have been implemented 
in the \Phantom Monte Carlo in a complete calculation with six fermions
in the final state.   
}

\begin{document}

\maketitle
        
\section{Introduction}
\label{sec:intro}

ATLAS \cite{ATLAS_HIGGS_2011} and CMS \cite{CMS_HIGGS_2011}
have recently presented the combination of all their Standard Model Higgs searches with the full
2011 dataset. Both experiments have registered a tantalizing excess of events at about 125 GeV with respect to the
hypothesis of the absence of the Higgs.
However the evidence is not yet sufficient to claim that
the Higgs boson has been detected and statistical fluctuations might well explain the data. 
Therefore,
the mechanism of ElectroWeak Symmetry Breaking (EWSB) remains unclear and   
the issue of high energy vector boson scattering continues to play a central role,
either as the final test of the nature of the Higgs boson or, if the Higgs doesn't show up,
as the main hunting ground for clues to alternative explanations.
The  vector vector scattering amplitudes grow with energy when the bosons are longitudinally polarized
and violate perturbative unitarity at about one TeV
\cite{Veltman:1976rt,Veltman:1977fy,Lee:1977yc,Lee:1977eg,Passarino:1985ax,Passarino:1990hk},
requiring either the Higgs or some new physics 
in the energy range accessible to the LHC in order to tame this unphysical
behaviour\footnote{ Detailed reviews and extensive bibliographies can be found in
Refs.~\cite{Djouadi:2005gi,Aad:2009wy,Ball:2007zza,Chanowitz:1998wi}}.

Many alternative mechanisms of EWSB  have been explored. We will not try to
summarize the different models and simply refer to the literature. 
These theories typically predict the presence
of new states which, much like the Higgs boson does, keep the scattering amplitudes
small and the full theory amenable to a perturbative treatment.
These additional particles, if light enough, could be observed at the LHC.
If no such state is present, the scattering amplitudes become strong as the energy 
increases and perturbation theory breaks down.
Moving beyond the perturbative approach it is possible to use Effective
Field Theory (EFT) methods, in particular the Electroweak Chiral Lagrangian (EWChL)
\cite{Appelquist:1980vg,Longhitano:1980iz,Longhitano:1980tm,Appelquist:1993ka,Contino:2006nn,Giudice:2007fh,Barbieri:2007bh},
to describe the low energy behaviour of the scattering amplitudes.
The EWChL is a powerful approach for treating the
low energy dynamics of systems with broken symmetries. It provides a systematic
expansion of the full unknown Lagrangian in terms of the fields which are
relevant at scales much lower than the symmetry breaking scale and does not require a detailed
knowledge of the full theory. 
It is then possible to apply Unitarization Methods,
using the lowest order terms in the scattering amplitudes
as building blocks of all order expressions which
respect unitarity and agree up to a finite order with the perturbative result.

Scattering processes among vector bosons have been scrutinized since a long time
\cite{Duncan:1985vj,Dicus:1986jg,Cahn:1986zv,Barger:1987du,Kleiss:1987cj,Barger:1988mr,Barger:1990py,Baur:1990xe,
Dicus:1990fz,Barger:1991ar,Dicus:1991im,Bagger:1993zf,Barger:1994zq,Bagger:1995mk,Rainwater:1999sd}
with an increasing degree of sophistication.
In a series of papers 
\cite{Accomando:2005hz,Ballestrero:2008gf,Ballestrero:2009vw,Ballestrero:2010vp} we have analyzed
all vector vector scattering channels,
$4j\ell\nu$, $4j\ell^+\ell^-$, $2j3\ell\nu$, $2j\ell^+\ell^-\nu\bar{\nu}$,
$2j\ell^\pm\ell^\pm \nu\nu$ and $2j4\ell$ ($\ell=e,\mu$)
which are observable at the LHC,
including all processes at order $\ordEW$, $\ordQCD$ and $\ordQCDsq$
as well as $t\bar{t}$+jets when appropriate.
We have systematically compared  a typical SM light Higgs scenario with the
Higgsless case and with an example of 
Strongly Interacting Light Higgs (SILH) model \cite{Giudice:2007fh}.

In the last few years QCD
corrections to boson--boson production via vector boson fusion
\cite{Jager:2006zc,Jager:2006cp,Bozzi:2007ur,Jager:2009xx}
at the LHC have been computed and turn out to be below 10\%. Recently, VBFNLO \cite{Arnold:2008rz},
a Monte Carlo program for vector boson fusion, double and triple vector boson production
at NLO QCD accuracy, limited to the leptonic decays of vector bosons, has been
released.

The first results for the NLO corrections to $W+4j$ production
have started to appear \cite{Berger:2010zx}.
New techniques which exploit the angular distribution of vector boson decay products
to determine the ratio of longitudinal and transverse polarization have been proposed
in \cite{Han:2009em}.

A number of papers
\cite{Hikasa:1991tw,Chanowitz:1993zh,Dobado:1999xb,Oller:1999me,Butterworth:2002tt,Chanowitz:2004gk,Englert:2008tn,Falkowski:2011ua}
have explored the possibility that the increasing strength of the interaction between vector bosons
as the energy of the scattering is augmented might lead to the formation of resonant states in analogy to Low Energy QCD
in which the growth of the pion--pion scattering amplitude is regulated by the appearance of the $\rho$ resonance.
Most of these efforts have resorted to the EVBA, since strictly speaking the unitarization procedure is defined only for on--shell
scattering amplitudes between longitudinally polarized vector bosons.
This approach however suffers of all the well known deficiencies of the EVBA and in particular neglects the contribution of
transversely polarized vector bosons and all off--shell effects which can be sizable.
The strong gauge cancellations between signal and irreducible background
and the reliability of the Equivalent Vector Boson Approximation (EVBA)
have been studied in  \cite{Accomando:2006vj}.
A preliminary analysis in the EVBA of the observability
of partial unitarization of longitudinal vector boson scattering in SILH models
at the LHC can be found in Ref.~\cite{Cheung:2008zh}.
Unitarization models can describe the strongly interacting symmetry
breaking sector, with or without formation of resonances.
A recipe for embedding the unitarized amplitudes in the full,
off--shell, $2 \rightarrow 6$ amplitude, which is necessary for a
reliable description of VV scattering at the LHC, has been first proposed in \rfs{Chanowitz:1995tn,Chanowitz:1996si}
and later refined in \rf{Alboteanu:2008my} and
made available in \Whizard \cite{Kilian:2007gr}.

In this paper we present the implementation of several unitarization schemes,
the K--matrix, the Inverse Amplitude Method (IAM) and the 
N/D procedure,
within the \Phantom \cite{Ballestrero:2007xq}
Monte Carlo event generator along the lines of \rf{Alboteanu:2008my}.
The unitarized amplitudes are extended off--shell in a natural way and embedded in the
framework of a complete six parton final state calculation.
We then study the prospect of detecting signals of these unitarized models at the LHC
taking into account all relevant backgrounds, including top--antitop production,
possibly with additional jets, and vector boson plus four jets production. 

The paper is organized as follows. In \sect{sec:implementation} we present the basic formalism of 
the EFT approach and we introduce the unitarization schemes
which are available in \Phantom.
In \sect{sec:comparison} we compare the results obtained in a complete calculation with those
obtained in the EVBA.
Then, in \sect{sec:counting} we present the main results for all relevant decay
modes of the final state bosons and finally we state our conclusions.

\section{Implementation of unitarized models in \Phantom}
\label{sec:implementation}

Pion-pion scattering has been for a long time described by effective Lagrangians.
In this framework, when the energy increases, non perturbative effects  of QCD dynamics are unavoidable
as for example manifested in the appearance of new resonances.
A purely phenomenological approach to describe this physics is given by Unitarization Models.
They are based on \emph{ad-hoc} formulas that force the amplitudes of Goldstone bosons scattering
to satisfy the unitarity condition and maintain the low energy behavior.
Unitarization models are intended to represent the approximate magnitude of these amplitudes
beyond the low-energy regime and have been able in some cases to describe some resonances of QCD.

In view of the great similarities between low energy QCD and Electroweak physics,
the ideas of unitarization models have been translated to a strong symmetry breaking sector in several studies
\cite{Bagger:1993zf,Alboteanu:2008my,Butterworth:2002tt,Chanowitz:1995tn,Chanowitz:1996si,Hikasa:1991tw,Dobado:1999xb}. 
They are not complete quantum field theories and in particular they typically violate
crossing symmetry, but despite these deficiencies, unitarization models fulfill
their phenomenological purpose of estimating the magnitude of strong $VV$ scattering
cross sections much beyond the range of validity of the effective theory.

In this section we describe the main aspects of the implementation of unitarization models
in a complete six-fermion in the final state framework for the \Phantom Monte Carlo generator.

\subsection{The Electroweak Chiral Lagrangian And Low Energy Amplitudes}

If EWSB is driven by new strong dynamics at TeV scale, the EWChL
(inspired by the Chiral Lagrangian of QCD) provides the most economical description of
electroweak physics below this scale. The EWChL accounts for all particles of the SM apart from the Higgs boson.
The gauge symmetry $SU(2)_L \times U(1)_Y$ is maintained by explicitly introducing the three Goldstone bosons, 
$\omega^a(x)$ with $a=1,2,3$, gathered in an $SU(2)$ matrix field  
\begin{equation}
\Sigma(x)=\exp\left(\frac{i\sigma^a\omega^a(x)}{v}\right),
\end{equation}
where $\sigma^a$ are the Pauli matrices and $v\approx246\,\GeV$ is the decay constant of
the Goldstone boson that gives the right masses to vector bosons. $\Sigma$ transforms under $SU(2)_L \times U(1)_Y$ as
\be
\Sigma\ra U_L(x)\Sigma U_Y^\dagger(x),
\ee
\be
U_L(x)=\exp\big(\frac{i\beta^a(x)\sigma^a}{2}\big),
\qquad U_Y(x)=\exp\big(\frac{i\beta_Y(x)\sigma^3}{2}\big).
\ee

The familiar pieces of the chiral Lagrangian, that emerge for example from the $M_H\ra\infty$ limit of the SM, are:
\begin{align}
\label{eq:ewchl}
  \mathcal{L}= &  \frac{v^2}{4}\,\mbox{Tr}[{(D_\mu\Sigma)^\dagger (D^\mu\Sigma)}]
     -\frac{1}{4}G_{\mu\nu}^aG^{\mu\nu,a}
     -\frac{1}{4}W_{\mu\nu}^iW^{\mu\nu,i} 
     -\frac{1}{4}B_{\mu\nu}B^{\mu\nu} \nonumber \\
 & + i\,\bar{Q}_L \sla{D} Q_L + i\,\bar{Q}_R \sla{D} Q_R
   + i\,\bar{L}_L \sla{D} L_L + i\,\bar{L}_R \sla{D} L_R \nonumber \\
 & - (\bar{Q}_L \Sigma M_Q Q_R + \bar{L}_L \Sigma M_L L_R + \mbox{h.c.}).
\end{align}
The first term has the form of a non-linear sigma model, which is non-renormalizable in four dimensions;
therefore, a cut-off scale $\Lambda$ must be set. 
Deviations from the SM in the absence of the Higgs boson
can be parametrized in terms of a low energy expansion in $E/\Lambda$,
consisting of operators of increasing dimension.

Besides the usual spontaneous breaking pattern of the SM, 
$SU(2)_L \times U(1)_Y \ra U(1)_{em}$, experiment demands that the Higgs sector also approximately respects a larger, 
$SU(2)_L\times SU(2)_R$ symmetry, $\Sigma\rightarrow U_L\Sigma U_R^\dagger$, which is
spontaneously broken to the diagonal subgroup $SU(2)_D$ (where $U_L=U_R$) by 
the VEV $\langle \Sigma \rangle=1$.
Only two dimension-4 operators which respect
these symmetries are relevant for the study of Vector Boson Scattering (VBS),
they are:
\begin{align}
\label{eq:ewchlhighorder1}
\mc{L}_4&=\alpha_4\mbox{Tr}[V^\mu,V^\nu]^2 ,\\
\label{eq:ewchlhighorder2}
\mc{L}_5&=\alpha_5\mbox{Tr}[V_\mu,V^\mu]^2 ,
\end{align}
where $V_\mu\equiv (D_\mu\Sigma)\Sigma^\dagger$.

For the description of the elastic scattering of longitudinal bosons, we are going to make use
of the Goldstone Boson (GB) amplitudes, which can be translated into
the corresponding physical longitudinal boson scattering through the Goldstone Boson Equivalence Theorem (GBET)
\cite{Cornwall:1974km,Lee:1977eg,Chanowitz:1985hj,Gounaris:1986cr}.
The GBET states that an amplitude involving longitudinal vector bosons, $V_L$, is 
well approximated at high energy by the
corresponding amplitude obtained by replacing $V_L$ with the Goldstone bosons, $\omega$, in any
renormalizable $R_\xi$ gauge. Assuming isospin custodial symmetry,
all $2\ra 2$ Goldstone boson scattering processes can be described by a single
master amplitude, $A(s,t,u)$ which satisfies $A(s,t,u)=A(s,u,t)$.
It can be identified as the amplitude of the $\gb^+\gb^-\ra zz$ process.

The lowest order term of the master amplitude in the $E/\Lambda$ expansion,
identified as the Low Energy Theorem (LET)\cite{Weinberg:1966kf} and reproduced
by the non-linear sigma model term in the Lagrangian (first term in the r.h.s. of \eqn{eq:ewchl}), is given by
\begin{equation}
\label{eq:wwzz_tree}
  A^{(1)}(s,t,u) = \frac{s}{v^2}\,.
\end{equation}
At next-to-leading order in the $E/\Lambda$ expansion, we must include the higher-dimension operators, 
\eqns{eq:ewchlhighorder1}{eq:ewchlhighorder2}, and one-loop diagrams.
The NLO amplitude is given by \cite{Cheyette:1987jf}
\bea
\label{eq:A_NLO}
  A^{(2)}(s,t,u) &=& 4\alpha_4\frac{t^2+u^2}{v^4} + 8\alpha_5\frac{s^2}{v^4}+ 
    \frac{1}{16\pi^2}\left[ \frac{10 s^2 + 13  (t^2 + u^2 )}{18 v^4} + \frac{s^2}{2v^4}\ln\left(\frac{\mu^2}{-s}\right)\right.\nonumber\\  
           &+& \left. \frac{t(s+2t)}{6v^4}\ln\left(\frac{\mu^2}{-t}\right) + 
      \frac{u(s+2u)}{6v^4}\ln\left(\frac{\mu^2}{-u}\right)\right],
\eea
where $\mu$ is the renormalization scale. The infinities that appear when computing
one loop diagrams are absorbed by defining renormalized parameters in the
higher-dimension operators of the effective Lagrangian.
The terms proportional to factors of the weak coupling, $g$, which do not grow
asymptotically with $s$, will be recovered     in the complete $2\ra 6$ implementation.
For the moment, we just need the leading contribution from the longitudinal boson scattering.

The $\alpha_4$ and $\alpha_5$ parameters contribute to the T--parameter
\cite{Peskin:1990zt,Peskin:1991sw,Barbieri:2004qk} and therefore are
constrained by electroweak precision data \cite{Dawson:1994fa,Eboli:2006wa}.
Stronger limits are obtained if quadratically divergent terms are taken into account
\cite{vanderBij:1997ec}.
Arguments based on unitarity and causality could also constrain these parameters
\cite{Fabbrichesi:2007ad}.
In this case, the magnitude of $\alpha_4$ and $\alpha_5$ is required by data 
to be smaller than about $10^{-2}$.

\subsection{Unitarization Of Low Energy Amplitudes}

In the EWChL framework, when $E$ approaches $\Lambda$, the perturbative expansion
starts to loose its predictive power. Moreover, low energy amplitudes of longitudinal
vector boson scattering violate unitarity at much lower scales 
$E\approx 1.2$ TeV. To describe the magnitude of the cross section much beyond the
unitarity violation scale we can unitarize the low energy amplitudes.

In order to perform the unitarization procedure, it is convenient to expand
the master amplitudes into isospin, $I$, eigenamplitudes according to
\footnote{These expressions as well as those in \eqn{PhysAmp} can be derived using the Clebsh--Gordan coefficients
for the coupling of two $I=1$ representation, exploiting the scalar nature of the interaction Hamiltonian.
The pion states are defined as: $|\pi^+ \!\!>$ = $1/\sqrt{2} (|\pi_1 \!\!>+ i\,| \pi_2 \!\!> )$ = -$ |1,1 \!\!>$,
 $|\pi^0 \!\!>$ =$ |\pi_3 \!\!>$ = 
$|1,0 \!\!>$, $ |\pi^-  \!\!>$ = $1/\sqrt{2} (|\pi_1 \!\!> - i\,| \pi_2 \!\!> )$ = $ |1,-1 \!\!>$}

\begin{subequations}
\label{eq:isospinamps1}
\begin{eqnarray}
  A_0(s,t,u) &=& 3A(s,t,u) + A(t,s,u) + A(u,s,t),\\
  A_1(s,t,u) &=& A(t,s,u) - A(u,s,t),\\    
A_2(s,t,u) &=& A(t,s,u) + A(u,s,t).
\label{eq:isospinampslast}
\end{eqnarray}
\end{subequations}

The individual scattering amplitudes can be expressed in terms of the isospin eigenamplitudes as follows:
\begin{subequations}
\begin{align}
  \label{A-wwzz}
   A(\gb^+\gb^-\to zz) &= \frac13 A_0(s,t,u)-\frac13 A_2(s,t,u)\,,\\
  \label{A-wzwz}
   A(\gb^+z\to \gb^+z) &= \frac12 A_1(s,t,u)+\frac12 A_2(s,t,u)\,,\\
  \label{A-wwww}
   A(\gb^+\gb^-\to \gb^+\gb^-) &= \frac13 A_0(s,t,u) + \frac12 A_1(s,t,u) + \frac16 A_2(s,t,u)\,,\\
  \label{A-wwww1}
   A(\gb^+\gb^+\to \gb^+\gb^+) &=  A_2(s,t,u)\,,\\
  \label{A-zzzz}
   A(zz\to zz) &= \frac13 A_0(s,t,u)+\frac23 A_2(s,t,u)\,.
\end{align}
\label{PhysAmp}
\end{subequations}

Each isospin eigenamplitude is then expanded into partial waves of definite
angular momenta, $J$, according to 
\be
\label{eq:spintoiso}
A_{IJ}(s)=\frac{1}{2}\int_{-1}^1 \de\cos\theta\, P_J(cos\theta)A_I(s,t,u),
\ee
where $P_J(x)$ are the Legendre polynomials.

Notice for instance the behavior of the $I=0,\,J=0$ eigenamplitude:
\be
\label{eq:A00}
A_{00}(s) = 2\frac{s}{v^2}+ \left[\frac83\left(7\alpha_4(\mu) + 11\alpha_5(\mu)\right)+ \frac{1}{16\pi^2}
    \left(2\ln\left(\frac{\mu^2}{-s}\right)+\frac{7}{9}\ln\left(\frac{\mu^2}{s}\right) + \frac{11}{54}\right)\right]\frac{s^2}{v^4}.
\ee
The absorptive part of this amplitude is extracted by setting $\ln(-s)=\ln s -i\pi$,
when $s>0$ and it reproduces the perturbative unitarity relation
\be
\Imm A^{(2)}_{IJ}(s)=\frac{1}{32\pi} |A_{IJ}^{(1)}(s)|^2\,,
\ee
where $A^{(1)}(s)$ is the lowest order amplitude (the first term of the r.h.s. of \eqn{eq:A00})
and $A^{(2)}(s)$ is the NLO part given by the term proportional to $s^2/v^4$.
This amplitude presents, up to one loop, the correct singularity structure, crossing property and chiral symmetry.
Nonetheless, this is not sufficient to
guarantee a well behaved cross section at high energies such as those of the LHC.

Therefore, each individual isospin-spin eigenamplitude must be unitarized
over the whole spectrum employing a specific protocol.
We have implemented the three most popular models of unitarization: the K-matrix scheme (KM),
the Inverse Amplitude Method (IAM) and the $N/D$ protocol.
The main purpose of these schemes is to transform the isospin-spin eigenamplitudes, $A_{IJ}(s)$,
into new expressions, $\hat{A}_{IJ}(s)$, which simultaneously respect the unitarity condition,
\be
\label{eq:unitaritycondition}
\Imm \hat{A}_{IJ}(s)=\frac{1}{32\pi} |\hat{A}_{IJ}(s)|^2\,,
\ee
and have the appropriate low energy behavior,
\be
\hat{A}_{IJ}(s)\xrightarrow{s\ra 0} A_{IJ}(s)\,.
\ee

\subsubsection{K-Matrix Scheme}
The unitarized amplitude through the K-Matrix scheme is given by
\be
\label{eq:km}
\hat{A}^{KM}_{IJ}(s)=\frac{32\pi}{32\pi \Ree(1/A_{IJ}(s))-i}\,.
\ee
This expression exactly satisfies elastic unitarity.
As an example, let us take the pure Low Energy Theorem expression, $A_{00}(s)=2s/v^2$.
The KM procedure transform this expression into 
\be
\hat{A}^{KM}_{00}(s)=\frac{2s}{v^2}\frac{1}{(1-\frac{i}{16\pi v^2}s)}
    \xrightarrow{s\ra \infty} 32\pi i\,,
\ee
which instead of growing quadratically with energy, asymptotically approaches unitarity saturation.

We have also implemented in the code the possibility of incorporating the exchange of
heavy resonances of definite angular momentum, $J$, and isospin, $I$, unitarized by the K-Matrix scheme.
We have followed the prescription given in \cite{Alboteanu:2008my},
adding new degrees of freedom to incorporate resonances coupled to the Goldstone bosons.
This unitarization procedure normally reproduces resonances with $s$-dependent widths.

\subsubsection{$N/D$ Protocol}

Besides violating crossing symmetry, the KM unitarization procedure spoils the
singularity structure of the fixed-order amplitudes.
It fixes the absorptive part for $s>0$ guaranteeing  unitarity at arbitrary energy at
the cost of ruining the left-hand cut. In the $N/D$ protocol, unitarity is exactly
restored with the extra quality of improved analytical properties.
In the $N/D$ method, each partial wave amplitude is expressed as the quotient of two functions,
\be
\label{eq:A_ND}
\hat{A}^{N/D}_{IJ}(s)=\frac{N_{IJ}(s)}{D_{IJ}(s)}\,.
\ee
The denominator function, $D(s)$, contains the right hand cut (or unitarity cut)
while the left-hand cut is incorporated in the numerator function, $N(s)$. 

In \eqn{eq:A00} for instance the left hand cut first appears through the
term $(7/9)\ln(\mu^2/s)$ which acquires an imaginary part for $s < 0$.
To unitarize the amplitudes and simultaneously reproduce the left hand cut of \eqn{eq:A00},
at one loop precision, we follow the prescription given by \cite{Oller:1999me}.
We define 
\be
G(s)=\frac{1}{32\pi^2}\ln\left(-\frac{s}{M^2}\right),
\ee
where $M$ is a new free parameter. The unitarized amplitudes take the form
\begin{eqnarray}
N_{IJ}(s)&=&A_{IJ}^{(1)}(s)+A_{IJ}^{(2)}(s)+G(s)(A_{IJ}^{(1)}(s))^2\,; \\
D_{IJ}(s)&=& 1+G(s)N_{IJ}(s)\,.
\end{eqnarray}
The terms $A_{IJ}^{(1)}(s)$ are originated from the lowest order amplitude, \eqn{eq:wwzz_tree},
while the  $A_{IJ}^{(2)}(s)$ are originated from the NLO terms, \eqn{eq:A_NLO}.
$N_{IJ}(s)$ bears exclusively the left hand cut, which is kept at the unitarized amplitude, $\hat{A}^{N/D}_{IJ}(s)$,
at the one-loop level precision. For $s>0$, $N_{IJ}(s)$ is real and
the unitarity cut is completely provided by the denominator function, $D_{IJ}(s)$,
making $\hat{A}^{N/D}_{IJ}(s)=\frac{N_{IJ}(s)}{D_{IJ}(s)}$ to exactly respect the
unitarity condition, \eqn{eq:unitaritycondition}.

\subsubsection{Inverse Amplitude Method}

In the Inverse Amplitude Method (IAM), isospin-spin eigenamplitudes are unitarized by the following prescription:
\be
\label{eq:iamtransf}
\hat{A}^{IAM}_{IJ}(s)=\frac{A_{IJ}^{(1)}(s)}{1-A_{IJ}^{(2)}(s)/A_{IJ}^{(1)}(s)}\,.
\ee

It can be shown that \eqn{eq:iamtransf} is a special case of the $N/D$ method,
it maintains the proper analytical structure with the correct branch cuts.
In addition, for certain values of the chiral coefficients, $\alpha_4$ and $\alpha_5$,
the unitarized amplitudes, both by $N/D$ and IAM protocols, present poles that can be interpreted as
dynamically generated resonances. 

This method has been widely and succefully applied in pion-pion and pion-kaon scattering,
in many cases reproducing lightest resonances.

\subsection{Implementing Unitarized Models in Six Parton Final States within \Phantom }

No detailed phenomenological study has been performed in the context of unitarization models
in a complete calculation. In most cases, previous results have relied on the
Effective Vector Boson Approximation (EVBA), which is known to produce inaccurate
predictions for VBS at high energy.
With a complete $2\ra 6$ calculation, all diagrams are summed coherently,
interference terms and off-shell effects are completely accounted for.
The contribution of transversely polarized bosons
is also correctly included.
A detailed description of the implementation of K Matrix unitarized model 
in \Whizard and some examples of simulation with the full 6 fermion calculation is
given in \rf{Kilian:2007gr}.

It is important to notice that the LET part is already present in the complete tree-level calculation.
In order to avoid double counting it is necessary to define a correction to the on-shell amplitudes, given by
\be
\Delta A_{IJ}(s)=\hat{A}_{IJ}(s)-A_{IJ}^{(1)}(s)\,.
\ee
These corrections to the isospin-spin eigenamplitudes are translated back into isospin eigenamplitudes according to 
\be
\Delta A_I(s,t) = \sum_J (2J+1)\Delta A_{IJ}(s)P_J(\cos\theta)\,.
\ee
For our purposes, it is enough to truncate the series at $J=2$.
Finally, the isospin eigenamplitudes are translated into corrections to the individual scattering amplitudes according to
\eqn{PhysAmp}.

We have for example:
\begin{align} 
\label{eq:a-wwzz} 
\Delta A(\gb^+\gb^-\rightarrow zz) &=  
    8\left[\frac{v^4}{24s^2}\left(\Delta A_{00}(s)-\Delta A_{20}(s) \right)-\frac{5v^4}{12s^2}\left(\Delta A_{02}(s)-\Delta A_{22}(s)\right)\right]
            \frac{s^2}{v^4} \nonumber \\
  & +4\left[\frac{5v^4}{4s^2}\left(\Delta A_{02}(s)-
            \Delta A_{22}(s)\right)\right]\frac{t^2+u^2}{v^4}.
\end{align}

In order to introduce each of these on-shell elastic scattering amplitudes into the
complete $2\ra 6$ matrix elements, a method based on the definition of new quartic
vertexes has been suggested in \cite{Chanowitz:1995tn,Chanowitz:1996si} and further developed in \cite{Alboteanu:2008my}.
Our implementation of this approach in the \Phantom Monte Carlo generator is briefly explained in the following.

\subsubsection{New Quartic Vertexes}

In unitary gauge the quartic gauge interactions of massive vector bosons can be
extracted from the EWChL, \eqns{eq:ewchl}{eq:ewchlhighorder2}, with $\Sigma=1$:
\begin{align}
\label{eq:QGC}
  \mc{L}_{QGC} &=  g^2\cos^2\tw\left[g_1^{ZZ} Z^\mu Z^\nu W^-_\mu W^+_\nu
              - g_2^{ZZ}Z^\mu Z_\mu W^{-\nu} W^+_\nu\right]\nonumber\\
       &+ \frac{g^2}{2}\left[ g_1^{WW}W^{-\mu} W^{+\nu} W^-_\mu W^+_\nu-
        g_2^{WW}\left(W^{-\mu} W^+_\mu\right)^2\right] \nonumber \\
  &+\frac{g^2}{4\cos^4\tw}h^{ZZ}(Z^\mu Z_\mu)^2\,.
\end{align}
Considering only the $\mc{L}_4$ and $\mc{L}_5$ extra terms, \eqns{eq:ewchlhighorder1}{eq:ewchlhighorder2},
the values of the couplings are given by \cite{Kilian:2007gr}
\begin{align}
\label{eq:smchiralparam}
g_1^{ZZ}&=1+\frac{g^2}{\cos^4\tw}\alpha_4\,,
     &g_2^{ZZ}&=1-\frac{g^2}{\cos^4\tw}\alpha_5\,,\nonumber\\
     g_1^{WW}&=1+g^2\alpha_4\,, & g_2^{WW}&=1-g^2(\alpha_4+2\alpha_5)\,, \nonumber\\
h^{ZZ}&=g^2(\alpha_4+\alpha_5)\,. &&
\end{align}
In the SM, $g_i^{VV}=1$ and $h^{ZZ}=0$, where $i=1,\,2$ and $VV=WW,\,ZZ$.

%

As already mentioned, the unitarization procedures used herein violate crossing symmetry,
therefore, it is not possible to write a Lagrangian for the model and
to derive the corresponding vertexes in a general form. Instead, it is necessary to identify first
which kind of vector vector scattering processes are embedded in the complete $2 \rightarrow 6$ set
of Feynman diagrams. As a second step  the vertexes are appropriately modified.
The possible scattering reactions, 
$W^+W^-\ra W^+W^-,\, W^+W^-\ra ZZ,\,W^\pm W^\pm\ra W^\pm W^\pm$, can be
identified from the flow of fermion external momenta.

For each scattering type, the quartic vertex is modified in such a way to reproduce the full
on--shell scattering amplitudes.
They can be written in the form:
\begin{subequations}
 \label{eq:modvertexes}
\begin{align}
    V(Z^\mu Z^\nu W^{-\rho}W^{+\sigma})  
       & = \!-ig^2\cos^2\tw[2k^{ZW}_1g^{\mu\nu}g^{\rho\sigma}\!-
                   k^{ZW}_2g^{\mu\sigma}g^{\nu\rho}\!-
                   k^{ZW}_3g^{\mu\rho}g^{\nu\sigma}], \\
   V(W^{-\mu}W^{+\nu}W^{-\rho}W^{+\sigma}) & = ig^2[2k^{WW}_1g^{\mu\rho}g^{\nu\sigma}\!-
                k^{WW}_2g^{\mu\sigma}g^{\nu\rho}\!-k^{WW}_3g^{\mu\nu}g^{\rho\sigma}], \\
   V(Z^\mu Z^\nu Z^{\rho}Z^{\sigma}) & = ig^2\frac{2}{\cos^4\tw}[k^{ZZ}_1g^{\mu\nu}g^{\rho\sigma}+
                 k^{ZZ}_2g^{\mu\sigma}g^{\nu\rho}+k^{ZZ}_3g^{\mu\rho}g^{\nu\sigma}]\,,
\end{align}
\end{subequations}
where the $k_i$ coefficients are form factors
which depend on the particular vector vector scattering process.
For example for $ZZ\ra W^+W^-$ we have:
\begin{subequations}
\label{eq:newvertexes}
\begin{align}
 k^{ZW}_1 &= 1-\frac{g^2}{\cos^4\tw}\left[\frac{v^4}{24s^2}\left(\Delta A_{00}(s) - \Delta A_{20}(s)\right)
   -\frac{5v^4}{12s^2}\left(\Delta A_{02}(s)-\Delta A_{22}(s)\right)\right], \\
 k^{ZW}_2 = k^{ZW}_3 &= 1+
  \frac{g^2}{\cos^4\tw}\left[\frac{5v^4}{4s^2}\left(\Delta A_{02}(s) - \Delta A_{22}(s)\right)\right]. 
\end{align}
\end{subequations}
Notice that for on-shell VBS, these new vertexes reproduce the unitarized amplitudes
(e.g. \eqn{eq:a-wwzz}) at high energies. Since $\epsilon^\mu_L\xrightarrow{E\gg M_V} \frac{p^\mu}{M_V}$,
in the high energy limit the Mandelstam variables of the scattering 
can be reexpressed in terms of the vector longitudinal polarizations in the form:

\begin{subequations}
\label{eq:lorentzfactors}
\begin{align}
        s^2&=2p_1\cdot p_2 \;2p_3\cdot p_4 \approx 4M_{V_1}M_{V_2}M_{V_3}M_{V_4}\;g_{\mu\nu}g_{\rho\sigma}\;
  \epsilon_1^\mu\epsilon_2^\nu\epsilon_3^\rho\epsilon_4^\sigma ,\\
  t^2&= 2p_1\cdot p_3 \;2p_2\cdot p_4 \approx 4M_{V_1}M_{V_2}M_{V_3}M_{V_4}\;g_{\mu\rho}g_{\nu\sigma}\;
  \epsilon_1^\mu\epsilon_2^\nu\epsilon_4^\rho\epsilon_4^\sigma ,\\
  u^2&=2p_1\cdot p_4 \;2p_2\cdot p_3 \approx  4M_{V_1}M_{V_2}M_{V_3}M_{V_4}\;g_{\mu\sigma}g_{\nu\rho}\;
  \epsilon_1^\mu\epsilon_2^\nu\epsilon_3^\rho\epsilon_4^\sigma , 
\end{align}
\end{subequations}
where the subindexes $1,\,2$ indicate the two incoming bosons and $3,\,4$ the two outgoing ones.
$M_V$ are the masses of the bosons, either $M_W$ or $M_Z$. With this identification,
the corrected amplitudes take exactly the form of quartic vertexes contracted with
external polarization vectors. The difference with respect to the complete six-fermions
final state calculation is that the role of the polarization vectors in \eqn{eq:lorentzfactors}
is played by the sub-diagrams with the final state fermions and the
boson propagators,
\be
\label{eq:polfork}
\raisebox{5ex}{$\epsilon^\mu$=}\,
\includegraphics[width=0.25\textwidth,height=0.12\textwidth]{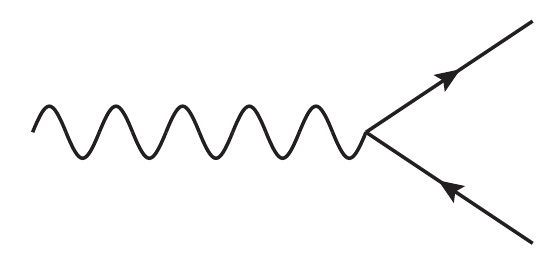}\,.
\ee

At low energies, the corrections $\Delta A_{IJ}$ tend to vanish; therefore,
the complete calculation is recovered. For high energies, the typical growth of
longitudinal vector boson scattering is moderated by the $\Delta A_{IJ}$ corrections
reproducing the unitarized elastic scattering of the GB, which is a good
approximation to the scattering of the longitudinal modes at high energy according to
the Goldstone Boson Equivalence Theorem.

\section{Comparison with EVBA results}
\label{sec:comparison}

The results of our implementation have been compared with those obtained in 
\rf{Butterworth:2002tt} by Butterworth, Cox and Forshaw, where  an on-shell calculation
based on the Effective $W$ Approximation has been  employed. 
Three instances of  the Inverse Amplitude Method, 
\eqn{eq:iamtransf}, have been considered. They differ in the choice 
of the $\alpha_4$ and $\alpha_5$ coefficients as reported in \tbn{tab:IAMscenarios}.
The renormalization scale has been set at  $\mu$=1 TeV. 

\begin{table}[h!tb]
\begin{center}
\begin{tabular}{|c|c|c|}
\hline
scenario 	& $\alpha_4(1\textrm{TeV})$ 	 & $\alpha_5(1\textrm{TeV})$ \\ 
\hline
\hline
A & $0.0$ & $0.003$ \\
\hline
B & $0.002$ & $-0.003$ \\
\hline
D & $0.008$ & $0.0$ \\
\hline
\end{tabular}
\caption{Parameters for the IAM models.}
\label{tab:IAMscenarios}
\end{center}
\end{table}

The IAM procedure gives rise to poles in the unitarized amplitudes  which can be interpreted as resonances whose mass and
width are given by  \eqn{eq:vectorparam} for the vector channel $I=1,\,J=1$ and by \eqn{eq:scalarparam}
for the scalar case $I=0,\,J=0$. Scenario A presents a scalar resonance at about 1 TeV,
scenario B a vector resonance at about 1.4 TeV, and scenario D a scalar resonance at 0.8 TeV and a vector one at 1.4 TeV. 

\be
\label{eq:vectorparam}
M_V^2=\frac{v^2}{4(\alpha_4-2\alpha_5)+\frac{1}{9(4\pi)^2}},\quad 
\Gamma_V=\frac{M_V^3}{96\pi v^2}.
\ee
\be
\label{eq:scalarparam}
M_S^2=\frac{12v^2}{16(11\alpha_5(M_S)+7\alpha_4(M_S))+\frac{101}{3(4\pi)^2}},\quad 
\Gamma_S=\frac{M_S^3}{16\pi v^2}.
\ee

\begin{table}[bth]
\begin{center}
\begin{tabular}{|c|}
\hline
\textbf{Generation cuts} \\
\hline
\hspace*{1cm} $p_T(\ell^\pm) > 20$ GeV \hspace*{1cm} \\
\hline
$|\eta(\ell^\pm)| < 3.0$ \\
\hline
$p_T(j) > 30$ GeV \\
\hline
$|\eta(j)| < 6.5$ \\
\hline
$M(jj) > 60$ GeV \\
\hline
$M(\ell^+\ell^-) > 20$ GeV \\
\hline
\end{tabular}
\caption{Standard acceptance cuts applied in the event generation and present
in all results of this section. Here $j = d,u,s,c,b,g$. } 
\label{tab:cuts_0}
\end{center}
\end{table}

\begin{figure}[hbt]
\begin{center}
\mbox{\includegraphics*[width=0.7\textwidth]{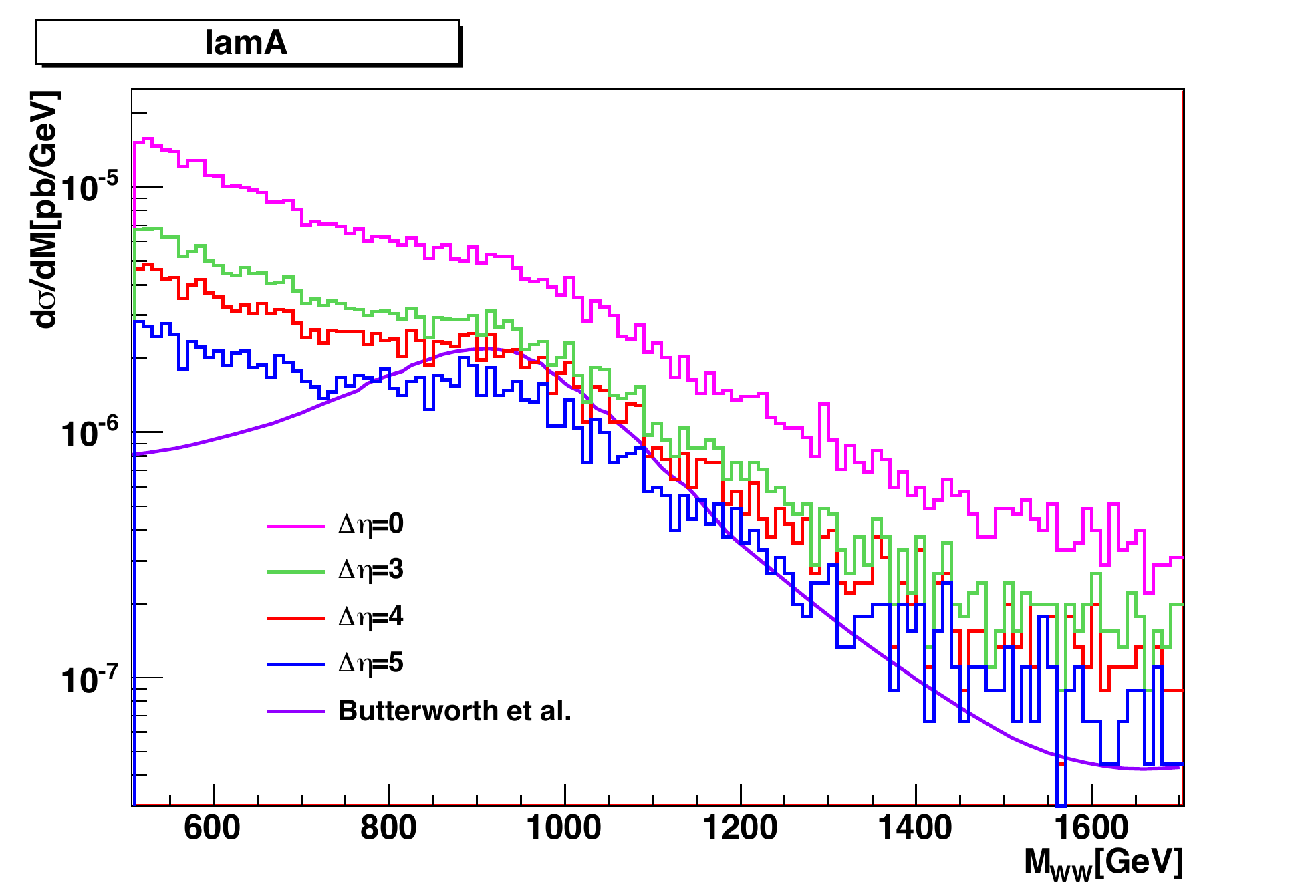}}
\caption{
Comparison between the $M(WW)$ distribution in $PP\ra 2j\mu^\pm e^\mp\nu\bar{\nu}$ in the complete
calculation and the EVBA calculation in \rf{Butterworth:2002tt} (in purple) multiplied
by the appropriate branching ratio, $BR=2/81$,
for the IAM A model.
The complete calculation is shown with different cuts in pseudo-rapidity.
Processes with external b-quarks have been discarded.
}
\label{fig:confr_iamA_cuts0345_Butt}
\end{center}
\end{figure}

We have computed the complete set of purely electroweak processes for the
$2j\mu^\pm e^\mp\nu\bar{\nu}$ final state, with the generation cuts in \tbn{tab:cuts_0}.
A big contribution to these final states comes from electroweak $t\bar{t}$ production.
Therefore, in order to avoid this contamination in the comparison,
we have considered only processes without bottom quarks in the initial or final state.
Moreover, different cuts in the pseudo-rapidity difference between tag jets have been
tested in order to enhance the scattering contribution.

\begin{figure}[h!bt]
\begin{center}
\mbox{\includegraphics*[width=0.5\textwidth,height=7cm]{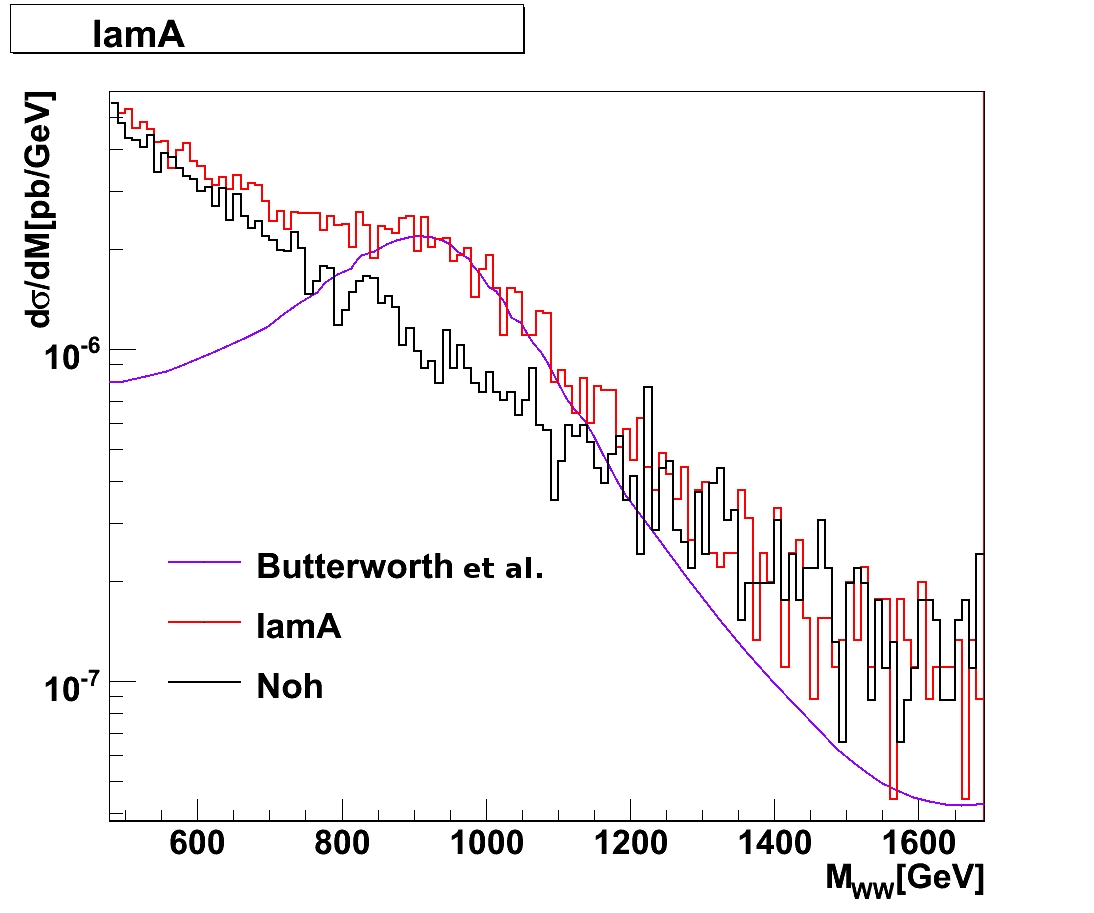}
      \includegraphics*[width=0.5\textwidth,height=7cm]{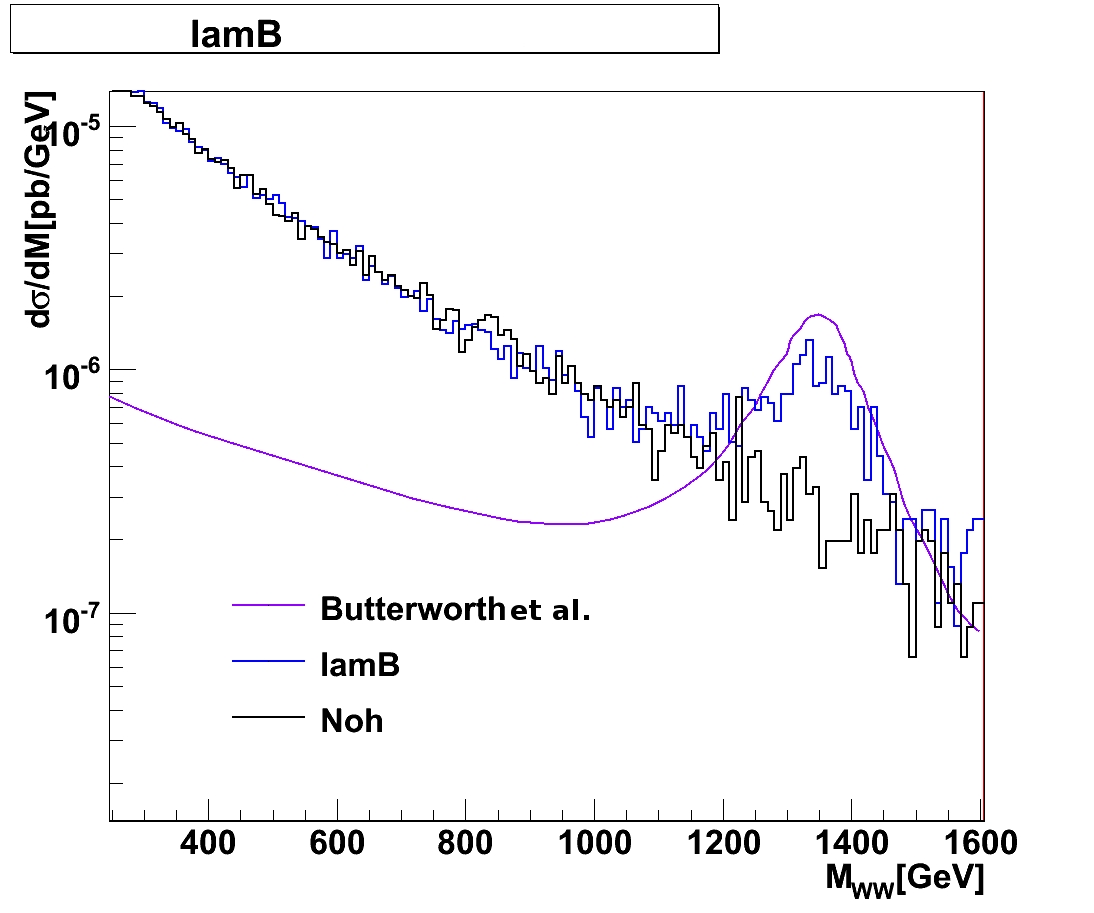}}
\mbox{\includegraphics*[width=0.5\textwidth,height=7cm]{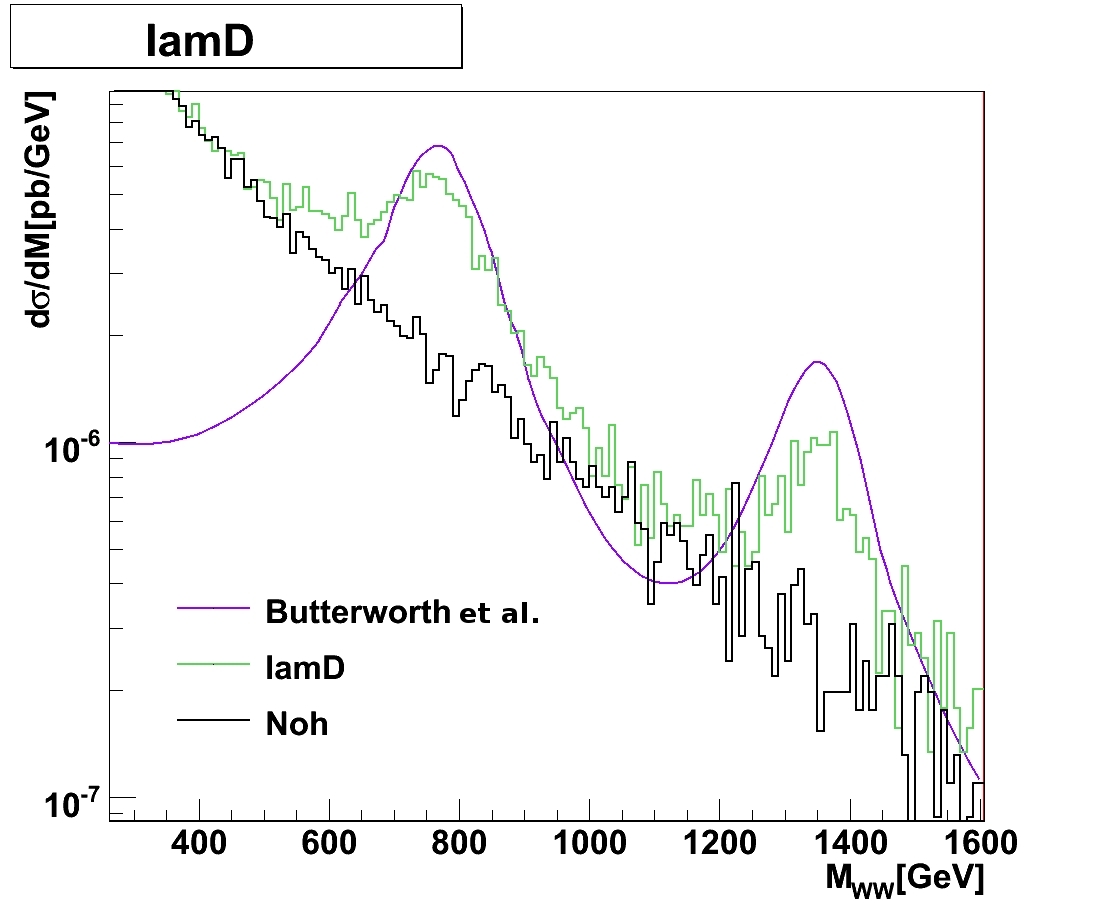}}
\caption{
Comparison between the  $M(WW)$ distribution in $PP\ra 2j\mu^\pm e^\mp\nu\bar{\nu}$ for the
IAM A (\emph{at top left}), IAM B (\emph{at top right}) and IAM D (\emph{bottom}) models obtained with a complete calculation
with $\Delta\eta>4$  and the predictions from 
\rf{Butterworth:2002tt} multiplied by the appropriate branching ratio, $BR=2/81$. 
}
\label{fig:IamABD_Butt}
\end{center}
\end{figure}

The $VV$ invariant mass distribution in the IAM A model is presented in \fig{fig:confr_iamA_cuts0345_Butt}.
It shows the prediction of the complete calculation, as implemented in \Phantom, and the
corresponding cross section derived in \rf{Butterworth:2002tt} in the EVBA.
The data points for the EVBA results have been extracted from the figures of \rf{Butterworth:2002tt} and rescaled
by the branching ratio for $ W^+W^-\ra \mu^\pm e^\mp\nu\bar{\nu}$,  $BR=2/81$, since the $W$ decays were not included.
For the complete calculation, four different cuts on the difference in pseudorapidity between jets,
$\Delta\eta(jj)>0,\,3,\,4,\,5$, have been applied.
As expected the resonance peak is more clearly seen at larger separation between tag jets.

It can be observed that the curve with $\Delta\eta>4$ is well reproduced by
the approximation in \rf{Butterworth:2002tt} close to the peak at about one TeV.
Outside this regions the curve of \rf{Butterworth:2002tt}  underestimate by a large amount the actual cross section.
The comparison between the present results and those obtained in EVBA should not be taken too literally.
In EVBA the final state jets have been completely integrated over and therefore no separation cut
between tag jets can be applied. While the approximate total cross section obtained is in rough
agreement with the full calculation in the peak region, too many details would be missing in a
simulation based on the EVBA to allow for a realistic study of the experimental observability
of these processes.

In \fig{fig:IamABD_Butt} we show the comparison for all three scenarios. As before, 
$\Delta\eta>4$ and no process with external b-quarks is included. Again, large discrepancies can be observed
outside the peak region. We also present in black the result for the no--Higgs scenario.
It is interesting to notice that the unitarized models
and the no-Higgs scenario are in reasonable agreement outside of the resonance peak in the considered mass range.
We will comment further on this point when dicussing the different final states in 
\sect{sec:counting}.

\section{Counting experiments for unitarized models at LHC}
\label{sec:counting}

In this section, a number of phenomenological studies at the LHC
for some typical choice of unitarization schemes and parameters are 
presented. We limit our analyses to the design energy of 
14 TeV and we do not attempt to determine the parameter range
for which resonances in the new scenarios
can be experimentally discovered. For some final states and for scalar 
resonances, looking for new resonances is just an extension of Standard Model
Higgs searches. For different kind of resonances however a completely different
strategy might be required.
We will therefore examine in which cases there will be a significant excess
of events at high invariant mass and if it will be possible to detect 
it at the LHC. For this reason we give to these searches the name of counting 
experiments.
All our results are at parton level only, and must be considered as a first
indication  of the effects of unitarized models in the context of a complete
lowest order calculation which avoids the
EVBA and takes all relevant irreducible backgrounds into
account.

In addition to the cases already considered in the previous
section (IAM A, IAM B and IAM D) we will consider the following scenarios:
\bd
\item{-} IAM C: it is another Inverse Amplitude Method unitarization model
        with parameters $\alpha_4=0.002$ and $\alpha_5=-0.001$. This model  
        produces a vector resonance at about 1.9 TeV. It was already discussed
        in \rf{Butterworth:2002tt}.
\item{-} N/D A: uses the N/D unitarization method with M = 1 TeV,
        $\alpha_4=0$ and $\alpha_5=0.003$. It presents a 
        1 TeV scalar resonance as in the IAM A model but with a softer resonance peak.
        Notice however that in general the N/D method gives models which can be
        quite different from the other methods we have considered.
\item{-} KM-LET and IAM E: these two scenarios contain no resonances at all. In the first case
        the Low Energy Theorem divergent contribution, neglecting all NLO terms,
        is unitarized using the  KM method. 
        In the second case the IAM procedure with $\alpha_4=0$ and $\alpha_5=0$ is employed.
        They will be compared to the no-Higgs scenario ($M_H \ra \infty$ in the Unitary Gauge SM)
        where no unitarization is performed.  
\ed

A detailed description of the comparison among all scenarios will be given for the
channel $2jW^+W^-\ra 2j\ell^+\ell^-\nu\bar{\nu}$, while for all other channels we will
limit ourselves to the IAM A, IAM B, IAM D, KM-LET and no-Higgs cases only.

We have generated about half a million unweighted events for each channel and scenario
and for the main backgrounds.  All $\ordEW$ and $\ordQCD$ processes have been included
in our definition of the signal.
At generation level the cuts of \tbn{tab:cuts_basic} have been used.
For each specific channel we have tried to determine which additional cuts could improve the
separation between the unitarized model results and those obtained in the SM with a light Higgs .
The specific cuts used at analysis level for each channel considered in the following are reported in
Appendix A. All results reported below are obtained after both basics and analysis cuts.

\begin{table}[ht!]
\begin{center}
\begin{tabular}{|l|}
\hline
{\bf Basic Cuts} \\
\hline
$p_T(\ell^\pm) > 20 \mbox{ GeV}$ \\
\hline
$|\eta(\ell^\pm)| < 3.0$ \\
\hline
$M(\ell^+\ell^-) > 20 \mbox{ GeV}$\\
$M(\ell^+\ell^-) > 250 \mbox{ GeV}$ \quad ($2jW^+W^-$)\\
$76$ GeV $< M(\ell^+\ell^-) < 106$ GeV \quad ($2jZZ$)\\
\hline
$p_T(j) > 30 \mbox{ GeV}$ \\
\hline
$|\eta(j)| < 6.5$ \\
\hline
$M(jj) > 60 \mbox{ GeV}$ \\
$M(j_fj_b)<70 \mbox{ GeV} ; M(j_fj_b)>100 \mbox{ GeV}$ \\
\hline
$|\Delta \eta (jj)| > 3.0$ \quad ($2j2\ell 2\nu$)\\
$|\Delta \eta (j_fj_b)| > 4.0$ \quad ($2j4\ell$, $4j\ell\nu$, $4j\ell\ell$)\\
\hline
$|M(jjj) - M_{top}| > 15 \mbox{ GeV}$ ($4j\ell\nu$, $4j\ell\ell$)\\
$|M(j\ell\nu_{rec}) - M_{top}| > 15 \mbox{ GeV}$ ($3\ell\nu + 2j$, $4j\ell\nu$)\\
\hline
$70 \mbox{ GeV} < M(j_cj_c) < 100 \mbox{ GeV}$  ($4j\ell\nu$, $4j\ell\ell$)\\
\hline
$\Delta R(jj) > 0.3$ ($4j\ell\nu$, $4j\ell\ell$)\\
\hline
\end{tabular}
\caption{Basic cuts used at generation level.} 
\label{tab:cuts_basic}
\end{center}
\end{table}


\subsection{$2jW^+W^-\ra 2j\ell^+\ell^-\nu\bar{\nu}$}
\label{subsec:2jWW}

The  $2j\ell^+\ell^-\nu\bar{\nu}$ final state is one of the most important channels
for the study of alternative symmetry breaking mechanisms at the LHC 
\cite{Ballestrero:2010vp}.
It is one of the channels in which the presence of two neutrinos in the final state
does not allow the reconstruction of the invariant mass of the boson pair produced
in association with two jets. This may appear as a limit since the invariant mass
of the two bosons represents the center of mass energy in the case of the
underlying boson boson scattering, and it is therefore the most appropriate variable 
to examine in order to discover signs of strong boson boson scattering. 
However, these final states are not
affected by the huge QCD background which is typical of 4 jets plus two leptons
final states. It has been shown  \cite{Bagger:1995mk,Ballestrero:2009vw} 
that using invariant mass distribution of
the two charged  lepton 
can give very good results in discriminating
between the light Higgs SM case and other scenarios.

As obvious, both 
 $2jW^+W^-\ra 2j\ell^+\ell^-\nu\bar{\nu}$ and
$2jZZ\ra 2j\ell^+\ell^-\nu\bar{\nu}$ contribute to this final state
when the flavour of the two final state leptons are the same.
However, cutting on the invariant $\ell^+\ell^-$ mass may easily discriminate
between the two channels.

We start considering the case of $2jW^+W^-\ra 2j\ell^+\ell^-\nu\bar{\nu}$
and we will use this channel to present a detailed analysis of the 
main features of the various unitarized models we consider.

\begin{figure}[h!]
\centering
\includegraphics*[width=\textwidth,height=6cm]{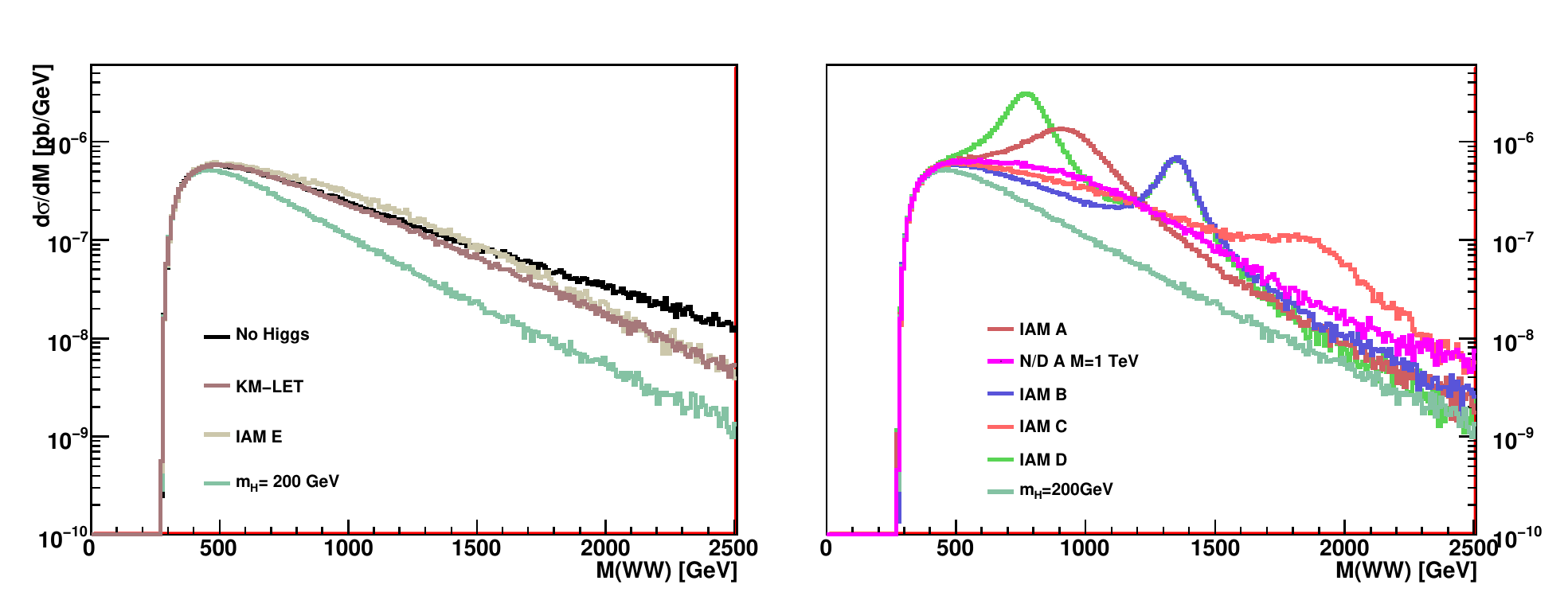}
\caption{WW  mass distributions in the  $2jW^+W^-\ra 2j\ell^+\ell^-\nu\bar{\nu}$
channel for different scenarios.}
\label{fig:ww_MWWsum}
\end{figure}

In \fig{fig:ww_MWWsum} one can see  the invariant mass distributions of the WW pair
in the different models.

These distributions cannot be measured in practice because the invariant
mass of two leptons and two neutrinos cannot be reconstructed. It is 
however possible to compute them at MC level and we present them
here in order to clearly see the behaviour of the various models.

On the left hand side of \fig{fig:ww_MWWsum} the distributions for the three
non resonant scenarios are reported and compared with the prediction of the SM
with a light Higgs.
Both unitarized distributions and the no-Higgs one are larger than the SM one and
the difference between strongly and weakly coupled theories can be easily
recognized. It is worth mentioning that the separation between the SM result and those
for the no--Higgs, KM--LET and IAM E cases is sharply increased by the additional
cuts in Appendix A
which enhance the boson boson scattering contribution.
The two unitarized models are very much similar, even if the unitarization scheme is 
different and in the IAM E case the NLO contributions are included.
The no-Higgs scenario is practically indistinguishable from
the previous ones up to an invariant mass of about 1.5 TeV.
Above this threshold the effects of the violation of unitarity in the
no-Higgs model manifest themselves. Because of this violation the no-Higgs model cannot be a 
consistent theory and can be valid only below 
an invariant mass of the order of a TeV. It has to be noticed however that 
at the LHC with design energy or lower, the vector vector pairs
produced in the bulk of the events will have a mass 
smaller than 1.5 TeV and as a consequence the violation of unitarity will be of small
practical relevance. Therefore the no-Higgs scenario can definitely be considered a 
good benchmark model to analyze the possibility of detecting new physics signals. 
This is  also evident from the right hand side plot in \fig{fig:ww_MWWsum}, where 
one can recognise the peaking structure of the various unitarized models we have 
considered. If the distribution could be actually  measured, the number of events of the various 
scenarios which develop resonances would be much larger than the ones of the left hand side.
It is also evident from the right hand side plot of this figure that both
scalar and vector resonances manifest themselves, as a consequence of the
fact that the boson boson scattering subdiagrams in this case correspond both to
$W^+W^-\ra W^+W^-$ (which produce both a vector and a scalar resonance) 
and to $ZZ\ra W^+W^-$ (which produce a scalar resonance).

\begin{figure}[h!]
\centering
\includegraphics*[width=\textwidth,height=6cm]{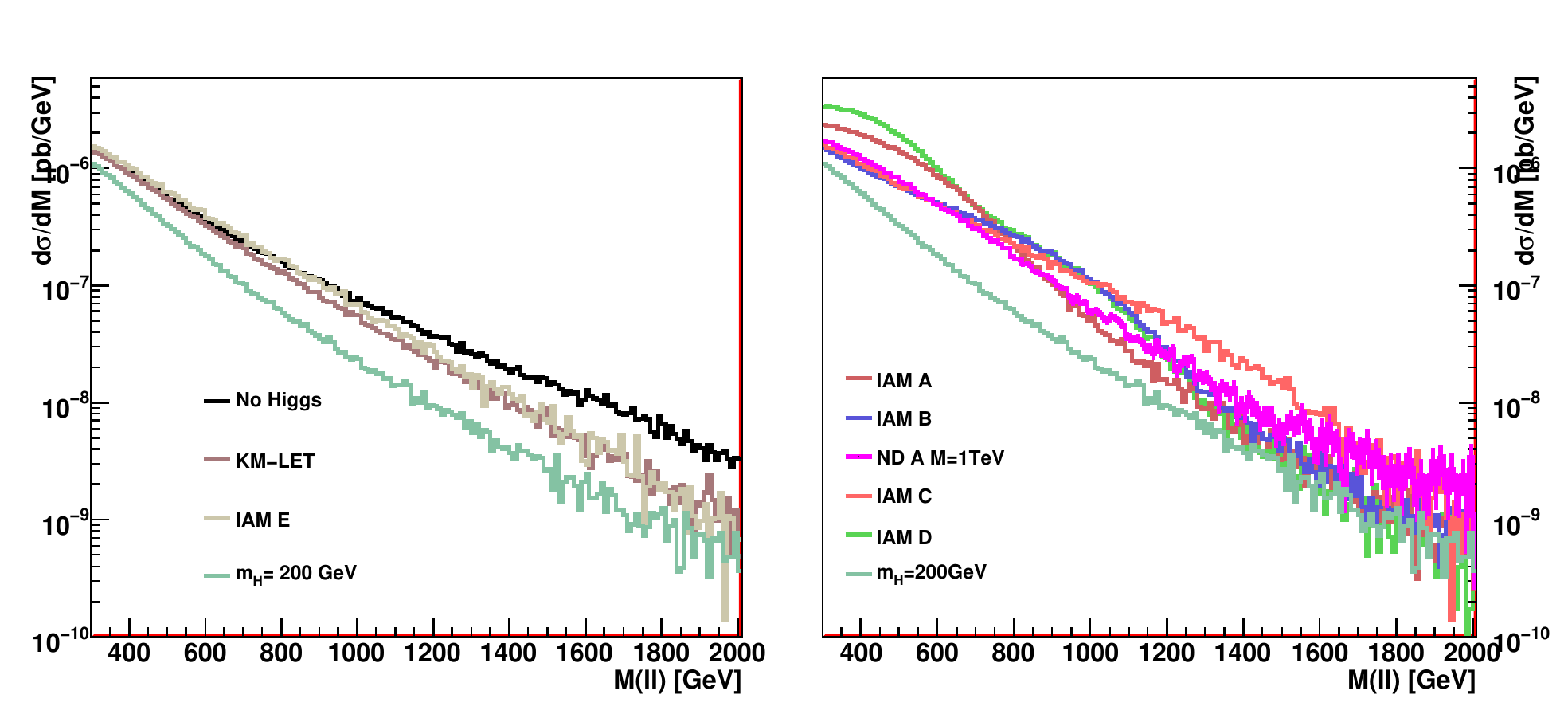}
\caption{Lepton-lepton mass distributions in the  $2jW^+W^-\ra 2j\ell^+\ell^-\nu\bar{\nu}$
channel for different scenarios.}
\label{fig:ww_Mllsum}
\end{figure}

In \fig{fig:ww_Mllsum} the distributions of the invariant mass of the two charged leptons
are presented. These distributions can be experimentally reconstructed
and can be employed to discriminate among the different physical models.
As it is obvious in these distributions all peaking structures have been 
smeared out but it is clear from the figures that the new physics models and 
the benchmark no-Higgs case can be clearly distinguished from a light Higgs scenario.
The difference between the light Higgs case and new physics grows with the $ll$
invariant mass for the non resonant models.
On the other hand the number of events at high invariant mass 
decreases. One has to find a good compromise in selecting a range of masses which
guarentees a good statistics and a sizeable difference between unitarization models and the SM.

This behaviour is exploited in \tbn{tab:res_ww}, where the cross sections are 
presented as a function of the minimun  $M(ll)$ invariant mass.
In the table the contribution of the $t\bar{t}jj$ background
which can be misinterpreted for the signal when the $b's$ go undetected
 is also reported \cite{Ballestrero:2010vp}.

\begin{table}[ht!]
\centering
\begin{tabular}{|c|c|c|c|c|c|c|c|c|c|c|}
\hline
$M_{cut}$ 	& no-H   & KM   & IAMA  &  IAMB	& IAMC  & IAMD   & IAME  & N/D   & SM   & $t\bar{t}jj$  \\ 
\hline
300             & .337  & .303  & .631  & .400  & .412  & .867   & .355  & .367  &.179  &.173 \\
400             & .212  & .186  & .413  & .274  & .277  & .547   & .224  & .240  &.100  &.0890\\
500             & .139  & .115  & .246  & .190  & .187  & .304   & .142  & .150  &.0577  &.0407\\
600             & .0968 & .0724 & .132  & .130  & .126  & .160   & .0897 & .0931 &.0332  &.0215\\
700             & .0696 & .0461 & .0658 & .0862 & .0858 & .0898  & .0571 & .0568 &.0217  &.0138\\
\hline
\end{tabular}
\caption{Total cross section (in fb) for the $(W^+W^-)\ell^+\ell^-\nu\bar{\nu}+2j$
channel with the full set of cuts in \tbn{tab:cuts_basic} and \tbn{tab:addcuts}
in function of the minimum $\ell\ell$ invariant mass, $M(\ell\ell)$ (in GeV).}
\label{tab:res_ww}
\end{table}

One can study the Probability Distribution Functions (PDFs) for the number of
events which can be measured at a given luminosity. This exercise can be repeated
for various $M_{cut}$ in order to find the optimum cut
for which the probability of dicriminating between the light Higgs case and new physics
scenarios is largest.
We show only the PDF's for the optimized cut. They are presented in 
\fig{PDFwwvv} for two reference luminosities $L=50\ifb$ and $L=200\ifb$. 
Here and in the following, the PDF's are computed assuming  Poissonian statistical 
fluctuations of the number of events computed by the MC and a theoretical
error on the number of signal events which we model as a flat distribution of
$\pm 30\%$ from the actual value  \cite{Ballestrero:2009vw,Ballestrero:2010vp}.

\begin{figure}[h!]
\centering
\subfigure{
\includegraphics*[width=0.52\textwidth,height=6cm]{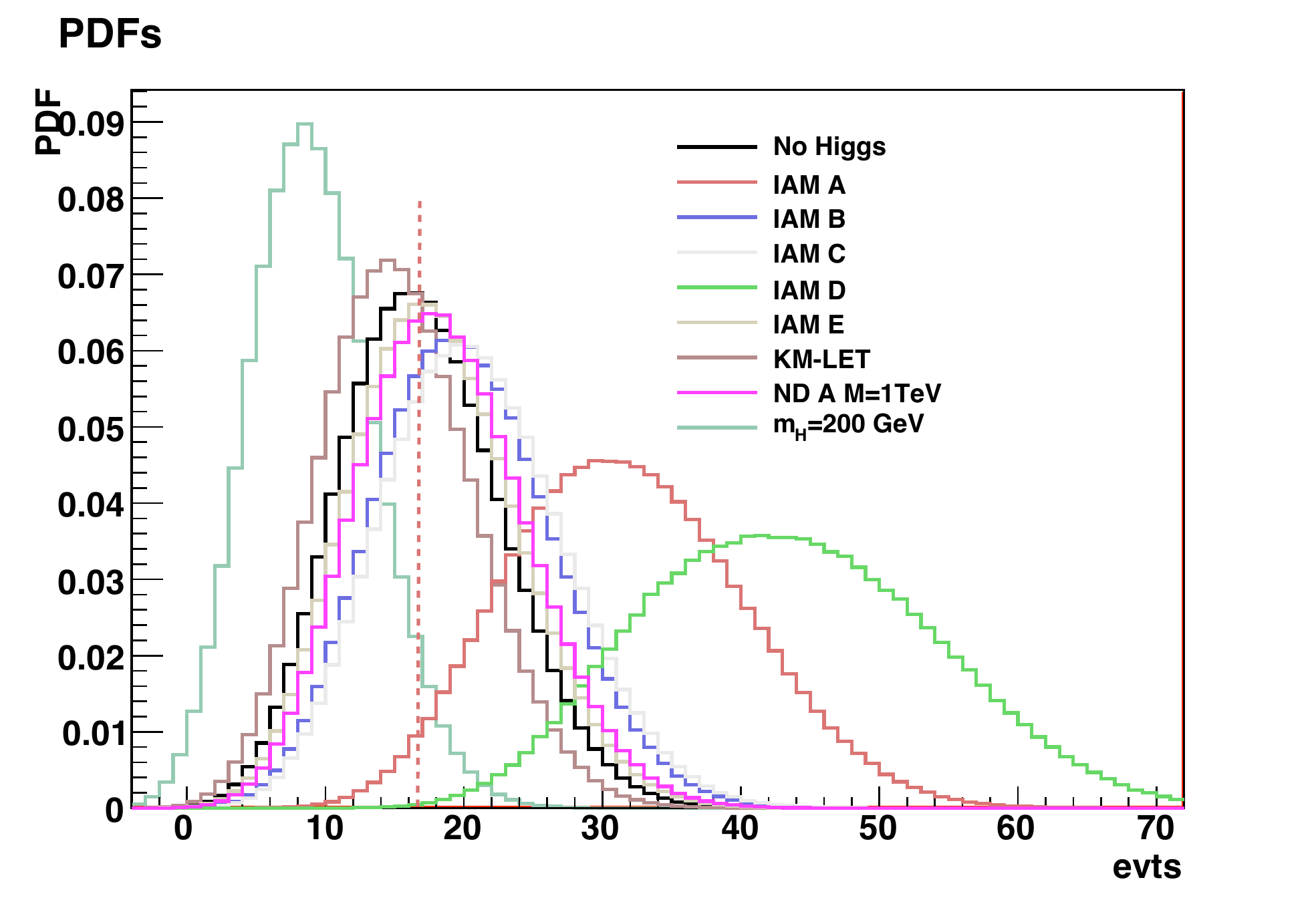}
\hspace*{-0.02\textwidth}
\includegraphics*[width=0.52\textwidth,height=6cm]{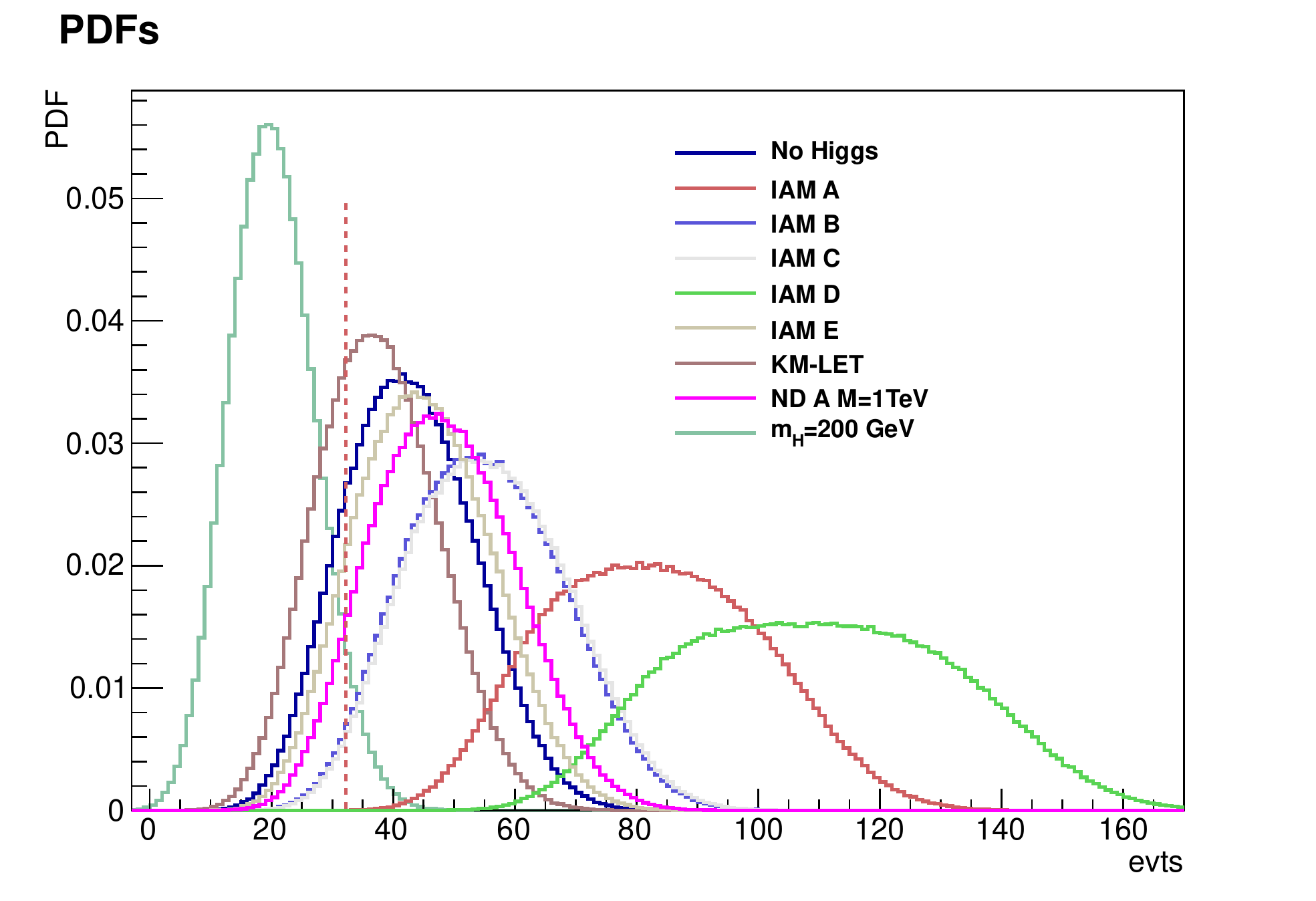}
}
\caption{PDFs of the number of events in the  $2jW^+W^-\ra 2j\ell^+\ell^-\nu\bar{\nu}$
channel for $M_{cut}=300$ GeV, $L=50\ifb$ (left) and $M_{cut}=400$ GeV, $L=200\ifb$ (right).}
\label{PDFwwvv}
\end{figure}

One can notice that the curves
corresponding to models with no resonances (KM-LET, no-Higgs and IAM E) are 
closest to 
the light Higgs distribution and are not much different from each other. The ones
which are most separated
are the ones corresponding to models with resonances, the separation being larger
for lower mass resonances.
In the plots the vertical line represents the 95\% limit of the
light Higgs distribution. We therefore compute what we call the PBSM@95\%CL
(Probability Beyond the SM at 95\% Confidence Level)
for the various scenarios as the probability that a number of events larger
than the 95\% limit occurs. The results, reported in \tbn{tab:pbsm_ww}, show quantitatively
that the probability of excluding the no-Higgs model
lies among the other non resonant ones.
Models with low mass resonances  can be easily excluded or verified already with luminosity $L=50\ifb$.
On the contrary a higher luminosity is required for models which are non resonant or
contain high mass resonances.

\begin{table}[ht!]
\centering
\begin{tabular}{|c|c|c|c|c|c|c|c|c|c|}
\hline
$L(\ifb)$ & $M_{cut}$ & no-H & KM      & IAMA  & IAMB     & IAMC     & IAMD     & IAME & N/D\\
\hline
   50   &   300	  & 49.4\%    & 37.8\%   & 97.2\% & 68.5 \%   & 71.6\%   & 99.9\%    & 55.4\%   & 59.2 \%\\	
  200   &   400    & 82.5\%    & 68.4  \% & 100\%  & 97.2 \%   & 97.5\%  & 100\%     & 87.2\%   & 91.8 \%\\
\hline
\end{tabular}
\caption{Probability to find a number of events larger than 95\% limit  of the SM 
(PBSM@95\%CL)  for the various scenarios in the  $2jW^+W^-\ra 2j\ell^+\ell^-\nu\bar{\nu}$
channel for two different luminosities $L=50\ifb$ and $L=200\ifb$. $M_{cut}$ is expressed in GeV.}
\label{tab:pbsm_ww}
\end{table}


\subsection{The other  two neutrinos final states}
\label{subsec:othervv}

The final state with two jets, two neutrinos and two charged leptons of different sign can result
from WW or ZZ production and decay. In a complete six fermion approach,
when the two leptons are of the same flavour, the separation of
the two channels has to rely on cuts on the $l^+l^-$ mass. Another well known and
interesting channel for studying boson boson scattering is  $2j\ell^\pm\ell^\pm\nu\nu$ in
which the two leptons are of the same sign. These two channels
will be discussed in the following in the context of unitarized models.

\subsubsection{The $2jZZ\ra 2j\ell^+\ell^-\nu\bar{\nu}$ channel}

As already mentioned, this channel has been separated from the
$2jWW \rightarrow 2j\ell^+\ell^-\nu\bar{\nu}$ with same flavour leptons
requiring $|M(\ell\ell)-M_Z|<15$ GeV.
A $ZZ$ final state can be produced in two different scattering processes:  $ZZ \ra ZZ$ 
and $W^+W^- \ra ZZ$.
As a consequence this final state has many analogies with the channel we have previously examined.
For such a reason we do not present
here the final state (not measurable) $ZZ$ invariant mass distribution. Instead we consider
the transverse $ZZ$ mass:
\begin{equation}
\label{eqn:mtzz}
M_T^2(ZZ)=[\sqrt{M_Z^2+p_T^2(\ell\ell)}+\sqrt{M_Z^2+p_{Tmiss}^2}]^2-|\vec{p_T}(\ell\ell)+\vec{p}_{Tmiss}|^2.
\end{equation}

The transverse mass distributions for the IAM A, IAM B, IAM D and KM-LET models
after basic and analysis cuts
are reported in \fig{fig:zz_MZZZZtsum} and compared with the no-Higgs and the light Higgs SM scenarios. 
It is rather evident that with the full set of cuts the non SM scenarios differ significantly from
the light Higgs case. It is in particular interesting that in this channel the IAM A and IAM D
scenarios are the two models which differ the most from the SM predictions.
They both present a relatively light scalar resonance.
On the contrary for the IAM B model the signal is not so prominent since
its vector resonance cannot appear in this channel.

\begin{figure}[h!]
\centering
\subfigure{	 
\includegraphics*[width=0.52\textwidth,height=6cm]{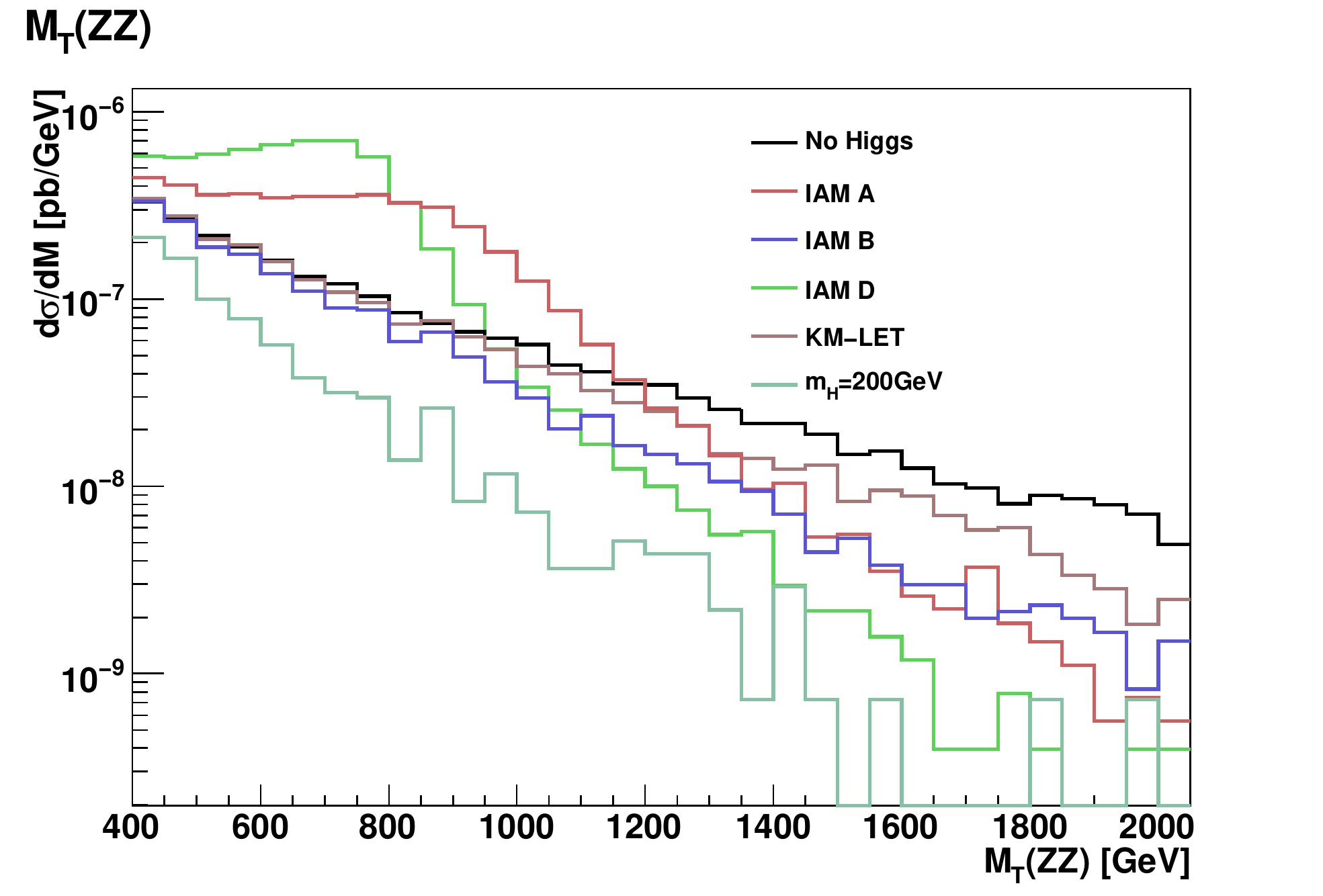}
}
\caption{Transverse mass distributions of the $ZZ$-system in the 
$2jZZ\ra 2j\ell^+\ell^-\nu\bar{\nu}$ channel
with full set of cuts, \tbn{tab:cuts_basic} and \tbn{tab:addcuts} of Appendix A.
}
\label{fig:zz_MZZZZtsum}
\end{figure}

Even if the transverse mass does not directly correspond to vector vector scattering center of mass energy,
the KM-LET and no-Higgs models have, as before, a similar 
behaviour for low invariant masses while they  start to deviate from each other
above 1 TeV. In terms of cross sections
their difference is only of the order of 10\% if we do not restrict ourselves
to very high invariant mass cuts.
The results for the cross sections  are reported in \tbn{tab:res_zz} .
The expected number of events is rather small.

\begin{table}[ht!]
\centering
\begin{tabular}{|c|c|c|c|c|c|c|}
\hline
$M_{cut}$      &no-Higgs& KM--LET	& IAM A	& IAM B	& IAM D	& SM \\ 
\hline
300             & .143  & .125  & .249  & .109  & .319  & .0540  \\
400             & .120  & .105  & .224  & .0886 & .291  & .0396 \\
500             & .0887 & .0739 & .181  & .0589 & .233  & .0214 \\
600             & .0691 & .0537 & .145  & .0408 & .172  & .0118 \\
700             & .0547 & .0395 & .110  & .0285 & .103  & .00697 \\
\hline
\end{tabular}
\caption{Total cross section (in fb) for the $(ZZ)\ell^+\ell^-\nu\bar{\nu}+2j$ channel 
with the full set of cuts in \tbn{tab:cuts_basic} and \tbn{tab:addcuts}
in function of the minimum $M_T(ZZ)$ (in GeV).
}
\label{tab:res_zz}
\end{table}

From the values reported in the \tbn{tab:res_zz}, we have determined the cut which 
maximizes the separation of the different scenarios which turns out to be $M_{cut}=500$ GeV
for $L=50\ifb$ and $M_{cut}=600$ GeV for $L=200\ifb$.
The corresponding Probability Distribution Functions are shown in \fig{fig:pbsm_zz} and confirm that
for the IAM A and IAM D models we are nearly certain to be able to observe a deviation from the SM
already with $50\ifb$. For all considered models, with $200\ifb$ of data,
the probability of observing a
discrepancy between the experimental results is above 80\%.

\begin{figure}[h!]
\centering
\subfigure{
\includegraphics*[width=0.52\textwidth,height=6cm]{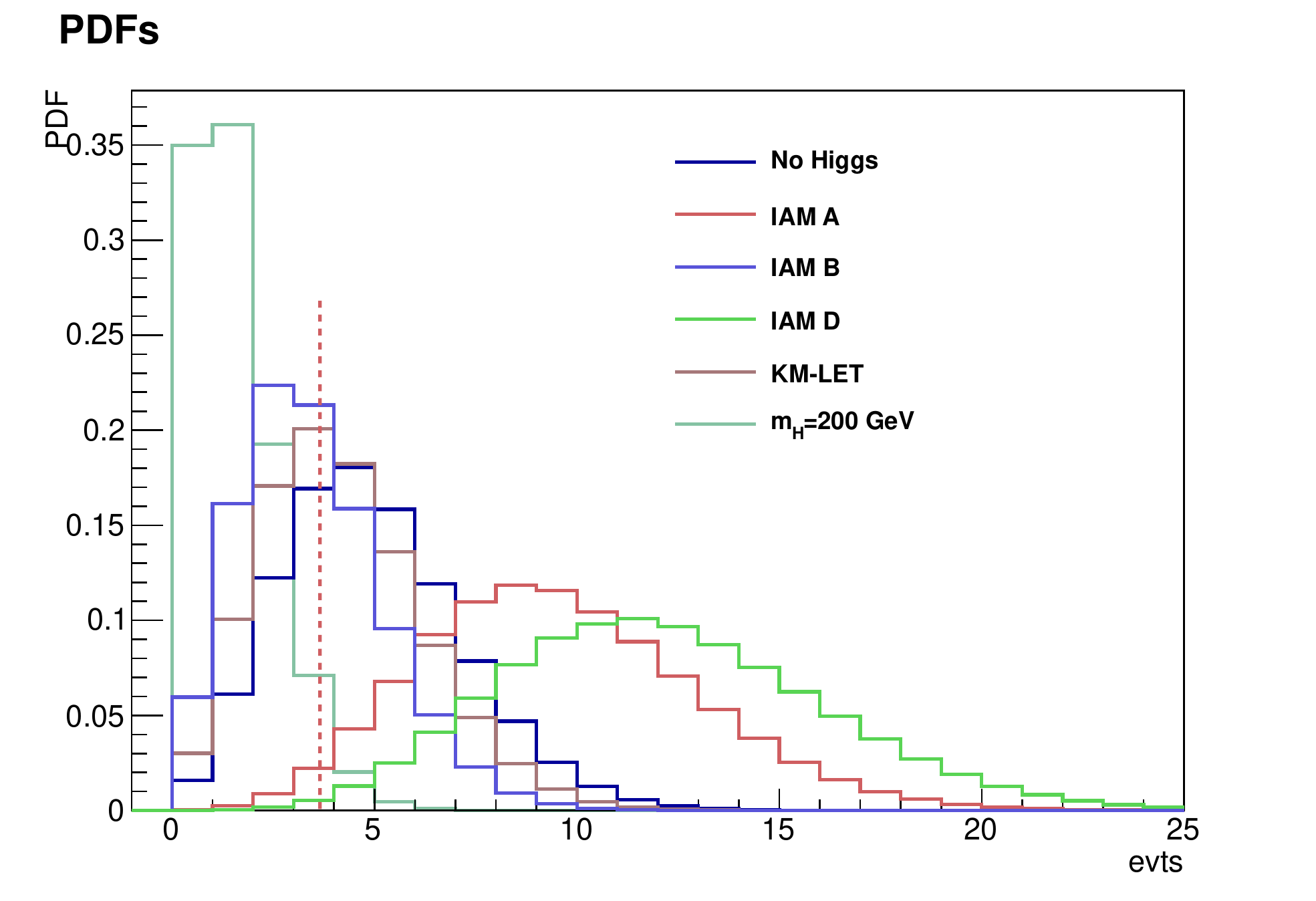}
\hspace*{-0.02\textwidth}
\includegraphics*[width=0.52\textwidth,height=6cm]{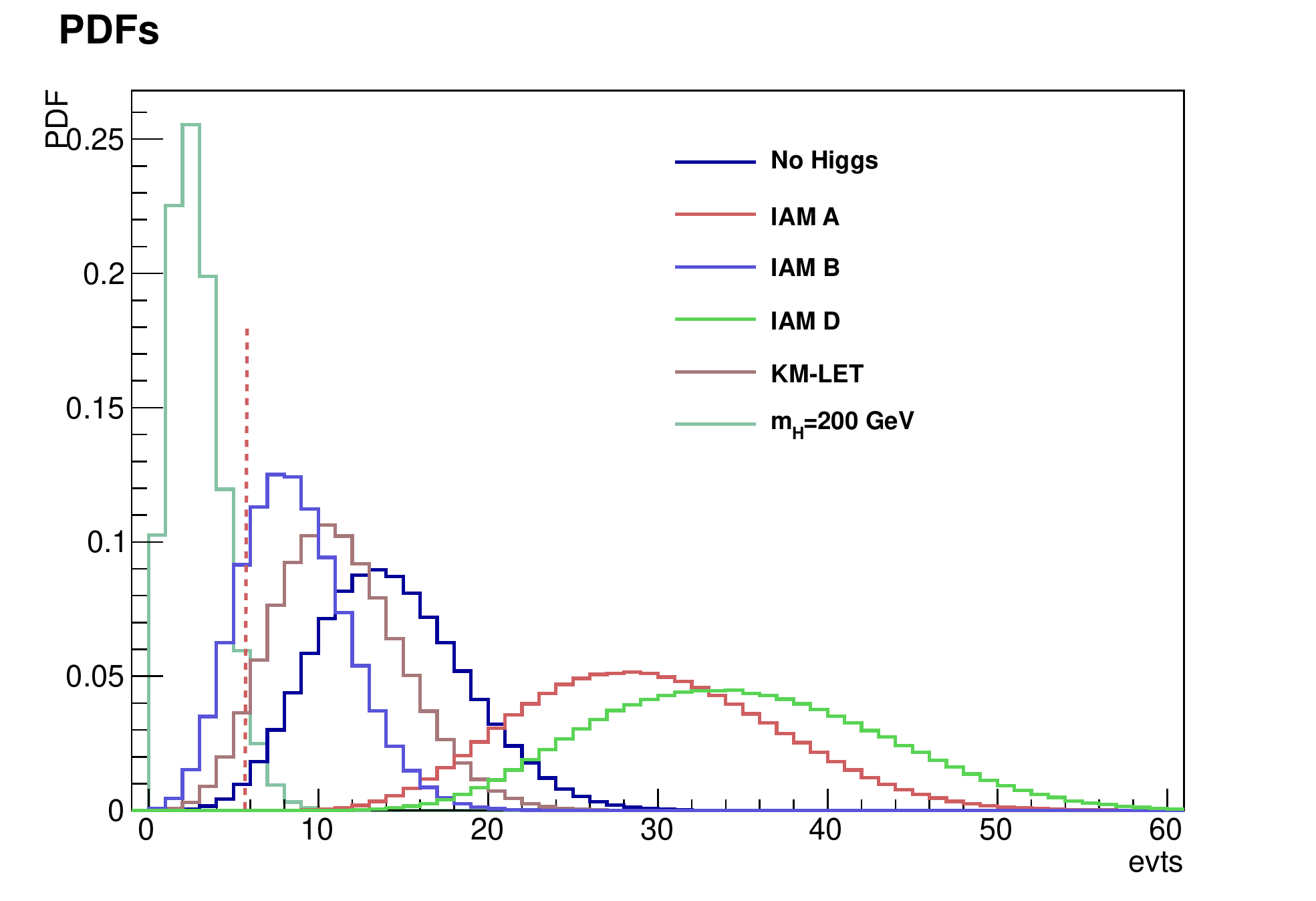}
}
\caption{PDFs of the number of events in the $(ZZ)\ell^+\ell^-\nu\bar{\nu}+2j$ channel
for $M_{cut}=500$ GeV, $L=50\ifb$ (left) and $M_{cut}=600$ GeV, $L=200\ifb$ (right).
}
\label{fig:pbsm_zz}
\end{figure}

If one compares \tbn{tab:pbsm_zz} with \tbn{tab:pbsm_ww}, it is apparent that in order
to discover signals of new physics in boson boson scattering processes, one must examine
all possible channels because some models can produce a modest signal in some channels
while being easily detectable in others.

\begin{table}[h!]
\centering
\begin{tabular}{|c|c|c|c|c|c|c|}
\hline
$L(\ifb)$ & $M_{cut}$ (GeV)	&  no--Higgs	& KM--LET	 & IAM A   	   & IAM B   	   & IAM D \\
\hline	
 50 & 500 & 68.8428\% & 56.5684\% & 97.3776\% & 41.4185\% & 99.4127\% \\
200 & 600 & 98.5618\% & 93.7842\% & 99.9998\% & 80.8113\% & 100\%\\
\hline
\end{tabular}
\caption{PBSM@95\%CL, defined in \subsect{subsec:2jWW}, for the various scenarios in the $2jZZ\ra 2j\ell^+\ell^-\nu\bar{\nu}$ 
channel.
}
\label{tab:pbsm_zz}
\end{table}

\subsubsection{The $2j\ell^\pm\ell^\pm\nu\nu$ channel}

The channel with two neutrinos and two same sign leptons in the final state
has long been considered one 
of the most promising because of the absence of large backgrounds. 
We will however see that the relevance of this channel depends quite a bit on which theory we are examining,
as this final state corresponds to an underlying scattering of two same sign $W$'s.

As usual we start considering the invariant mass distributions of the two leptons. 
They are shown in  \fig{fig:zz_MWWllsum}. The full set of 
cuts has been imposed. No peaking structure is present.
This reflects the fact that, for the models we have examined, neither scalar nor 
vector resonances can be produced in this channel. In this case the no-Higgs and 
KM-LET models, which do not develop resonances in any channel, give the largest predictions.
The lowest curve is as usual the SM light Higgs scenario.
All the others fall in between. 

\begin{figure}[h!]
\centering
\subfigure{	 
\includegraphics*[width=0.52\textwidth,height=6cm]{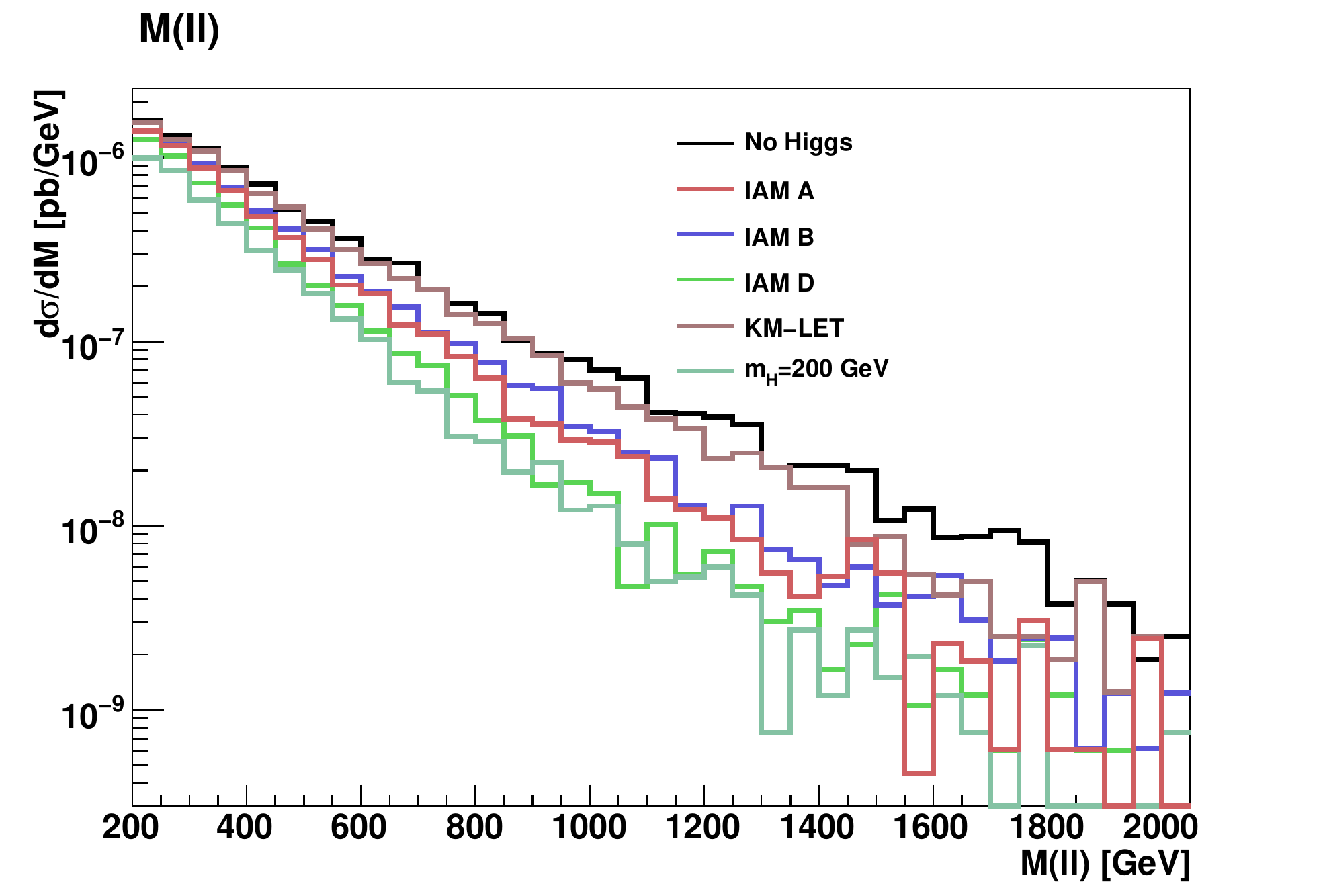}
}
\caption{Distributions  of the $\ell\ell$ invariant mass for the $2j\ell^\pm\ell^\pm\nu\nu$ channel
with full set of cuts, \tbn{tab:cuts_basic} and \tbn{tab:addcuts}.}
\label{fig:zz_MWWllsum}
\end{figure}

These results are made quantitative by the values of the cross sections as a function
of the dilepton invariant mass cut which is shown in \tbn{tab:res_llss} and by the
PDF distributions of \fig{fig:PDFsamesignw}.

\begin{table}[ht!]
\centering
\begin{tabular}{|c|c|c|c|c|c|c|}
\hline
$M_{cut}$      &no--Higgs& KM-LET & IAM A	& IAM B	& IAM D & SM \\ 
\hline
200		& .435  & .407  & .310  & .332  & .254  & .206	\\
300             & .290  & .267  & .183  & .201  & .141  & .114	  \\
400             & .191  & .171  & .107  & .120  & .0768 & .0629	   \\
500             & .129  & .112  & .0643 & .0744 & .0430 & .0351	   \\
600             & .0886 & .0760 & .0403 & .0474 & .0250 & .0194	   \\
700             & .0614 & .0517 & .0250 & .0304 & .0150 & .0112	   \\
\hline
\end{tabular}
\caption{Total cross section (in fb) for the $\ell^\pm\ell^\pm\nu\nu+2j$ channel 
        with the full set of cuts in \tbn{tab:cuts_basic} and \tbn{tab:addcuts}
in function of the minimum $\ell\ell$ invariant mass, $M(ll)$ (in GeV).
}
\label{tab:res_llss}
\end{table}

\begin{figure}[t!]
\centering
\subfigure{
\includegraphics*[width=0.52\textwidth,height=6cm]{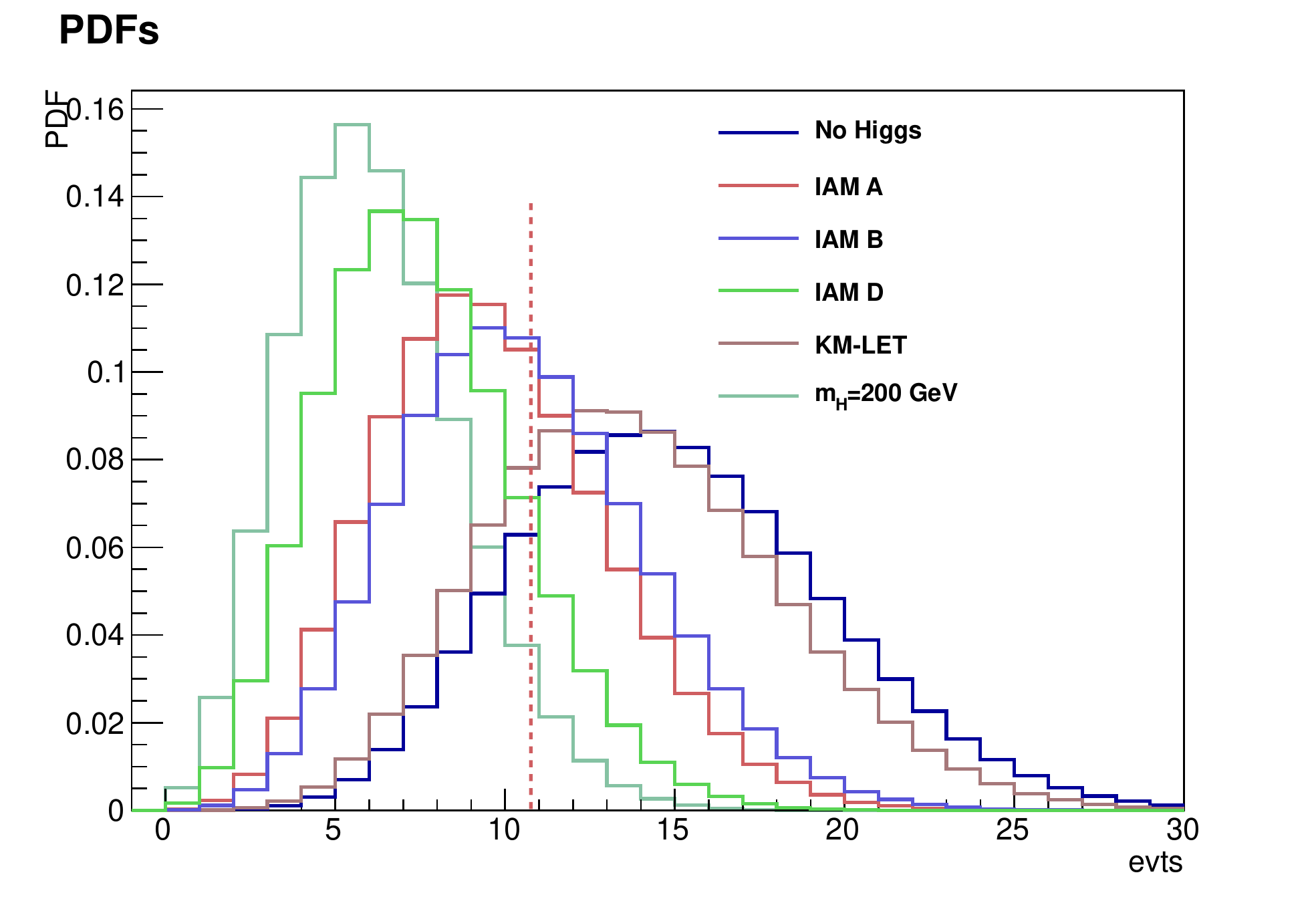}
\hspace*{-0.02\textwidth}
\includegraphics*[width=0.52\textwidth,height=6cm]{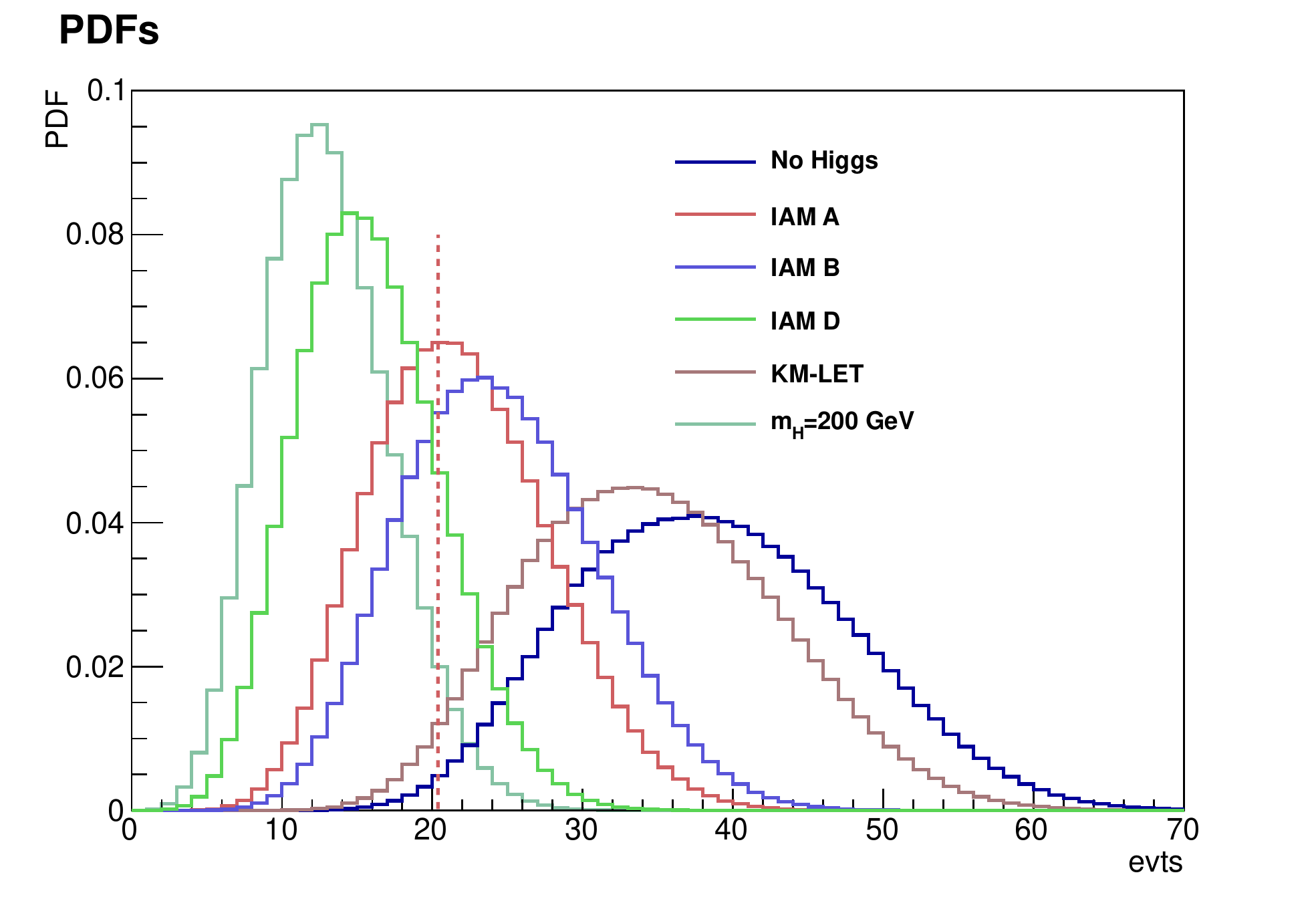}
}
\caption{PDFs of the number of events in the $\ell^\pm\ell^\pm\nu\nu+2j$ channel
for $M_{cut}=300$ GeV, $L=50\ifb$ and $M_{cut}=400$ GeV, $L=200\ifb$.}
\label{fig:PDFsamesignw}
\end{figure}

One sees immediately that the number of events 
produced in this channel is not large. Moreover the IAM A, IAM B and IAM D models
can hardly be distinguished from the light Higgs SM case. \tbn{tab:pbsm_llss} shows however that
non resonant unitarized models can give detectable signals of new
physics at high luminosity.

\begin{table}[h!]
\centering
\begin{tabular}{|c|c|c|c|c|c|c|}
\hline
$L(\ifb)$ & $M_{cut}$ (GeV) & no--Higgs& KM-LET & IAM A   & IAM B   & IAM D \\
\hline	
 50     & 300	   & 81.42\%  & 74.36\%  & 34.48\% & 44.36\% & 13.60\%\\
200     & 400      & 98.96\%  & 96.92\%  & 57.72\% & 72.03\% & 17.21\%\\
\hline
\end{tabular}
\caption{PBSM@95\%CL, defined in \subsect{subsec:2jWW}, for the $\ell^\pm\ell^\pm\nu\nu+2j$ channel.}
\label{tab:pbsm_llss}
\end{table}

In conclusion, one has to be very careful in relying on this channel for
detecting new physics because, even if it has the advantage of 
very small backgrounds, it is not sensitive to the types of resonances considered here. 
On the other hand it achieves one of the best dicriminating powers for scenarios without resonances.


\subsection{Final states in which the boson boson mass
can be reconstructed}
\label{subsec:reconstructed}

The final states in which the boson boson mass $M_{VV}$
can be reconstructed are those
in which at most one neutrino is produced, namely 
$\ell\nu+4j$, $\ell^+\ell^-+4j$, $3\ell\nu+2j$  and $4\ell+2j$. Reconstructing
$M_{VV}$ is very useful because it
corresponds to C.M. energy of the underlying boson boson scattering. As it
happens for on--shell boson boson scattering, the differences between
weakly and strongly interacting theories manifest themselves more clearly
at high invariant mass. 

The $4\ell+2j$ channel, with two opposite sign charged lepton pairs has a very low 
cross section and it corresponds to a ZZ final state.

In order to compute the invariant mass
of the two bosons, when a neutrino is present in the final state
its longitudinal momentum is reconstructed with the usual
procedure of forcing the 
invariant mass of the $\ell \nu$ pair to be equal to the $W$ boson nominal mass,
\begin{equation}
\label{eq:nu_reco_equation}
(p^{\ell}+p^{\nu})^2 = M_W^2.
\end{equation}
This determines, up to a two fold ambiguity, the longitudinal component of the neutrino 
momentum \cite{Ballestrero:2008gf}.

The $4j\ell\nu$ and  $\ell^+\ell^-+4j$ channels contain 4 final state jets.
Two of them, the most forward and most backward one, will provide tagging
and will be required to be well separated and energetic.
The two central ones will be
considered as candidates to reconstruct an electroweak boson. The
boson boson invariant mass will be assumed to be that of the system
formed by  the two central jets,
the charged lepton and the reconstructed neutrino.

The cross section of the four jet channels is relatively large, however they
are affected by substantial backgrounds. In particular $V + 4 j$ QCD processes give very 
large rates even after cut optimization. For the $4j\ell\nu$ final state also 
$t\bar{t}$ and $t\bar{t}+jets$ production has to be considered carefully. 
We have discussed in detail in our previous papers \cite{Ballestrero:2008gf,
Ballestrero:2009vw,Ballestrero:2010vp}
how to deal with these backgrounds. Here we only recall that when computing
the Probablity Distribution Functions we will attribute a statistical error
to these contributions but no theoretical error since we assume that they can be
measured in nearby regions and extrapolated to the kinematic range of interest for
VV scattering. 

\subsubsection{The $4j\ell\nu$ channel}

The final state of this cannel corresponds to both ZW and WW production.
The invariant mass distributions of the boson pair with the full set
of cuts is reported in \fig{fig:4jlv_MWV}. The corresponding
cross sections for various values of the invariant mass cut are given
in \tbn{tab:res_4jlv}. Both scalar and vector resonances are clearly
visible in this channel. The $W+4j$ background is much larger than the
SM electroweak prediction; it turns out to be of the same order as the expected yield
of the KM-LET and no-Higgs non resonant models.
It must be pointed out that this background, computed
at tree level with \MadEvent at a fixed scale, has been rescaled to the 
scale $Q=\hat{H}_T'/2$, $\hat{H}_T'=\sum_i^4 p_T(j) + E_T(W)$, 
where $E_T^2(W)=p_T^2(W)+M^2(W)$ \cite{Berger:2010zx}. With this dynamical scale the $W+4j$
contribution has sensibly decreased compared to the
fixed scale results.

\begin{figure}[h!]
\centering
\includegraphics*[width=0.62\textwidth,height=6cm]{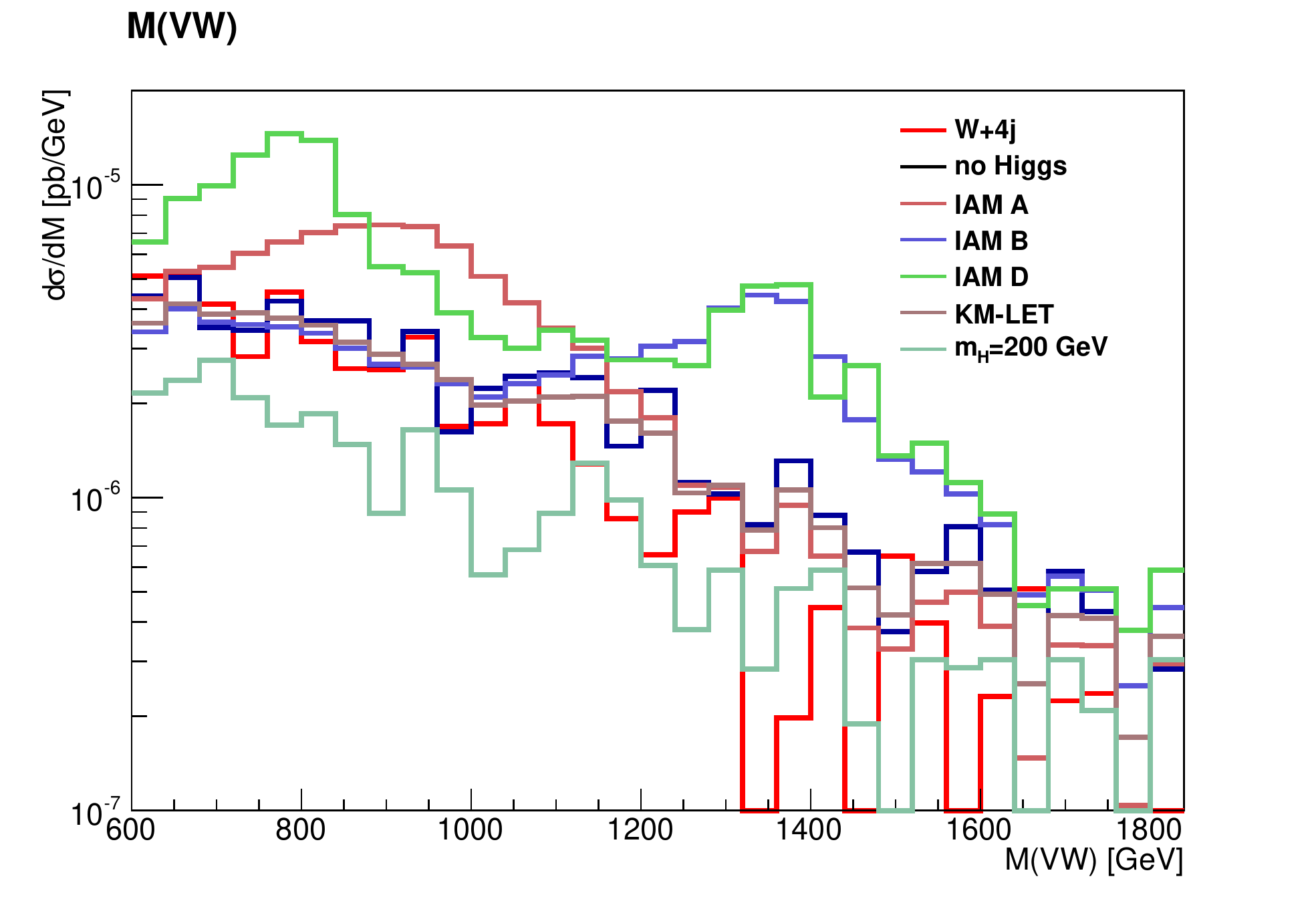}
\caption{Distributions of the mass  of the $VW$-system for the $\ell^\pm\nu+4j$ channel 
with full set of cuts, \tbn{tab:cuts_basic} and \tbn{tab:addcuts}.}
\label{fig:4jlv_MWV}
\end{figure}

\begin{table}[ht!]
\centering
\begin{tabular}{|c|c|c|c|c|c|c|c|c|}
\hline
$M_{cut}$      &no--Higgs& KM--LET & IAM A	& IAM B	& IAM D & SM   & $W+4j$ & $t\bar{t}+2j$\\ 
\hline
600		& 2.36  & 2.23  & 3.66  & 3.04  & 5.46  & 1.048 & 2.03   & .432\\
800             & 1.558 & 1.46  & 2.56  & 2.32  & 3.36  & .618  & 1.15   & .167\\
1000            & .966  & .877  & 1.13  & 1.76  & 1.90  & .338  & 0.62   & .0617\\
1200            & .526  & .478  & .414  & 1.26  & 1.27  & .188  & 0.30   & .0264\\
\hline
\end{tabular}
\caption{Total cross section (in fb) for the $\ell^\pm\nu+4j$ channel 
with the full set of cuts in \tbn{tab:cuts_basic} and \tbn{tab:addcuts}
in function of the minimum $j_cj_c\ell\nu$ invariant mass (in GeV).
}
\label{tab:res_4jlv}
\end{table}

\begin{figure}[t!]
\centering
\subfigure{
\includegraphics*[width=0.52\textwidth,height=6cm]{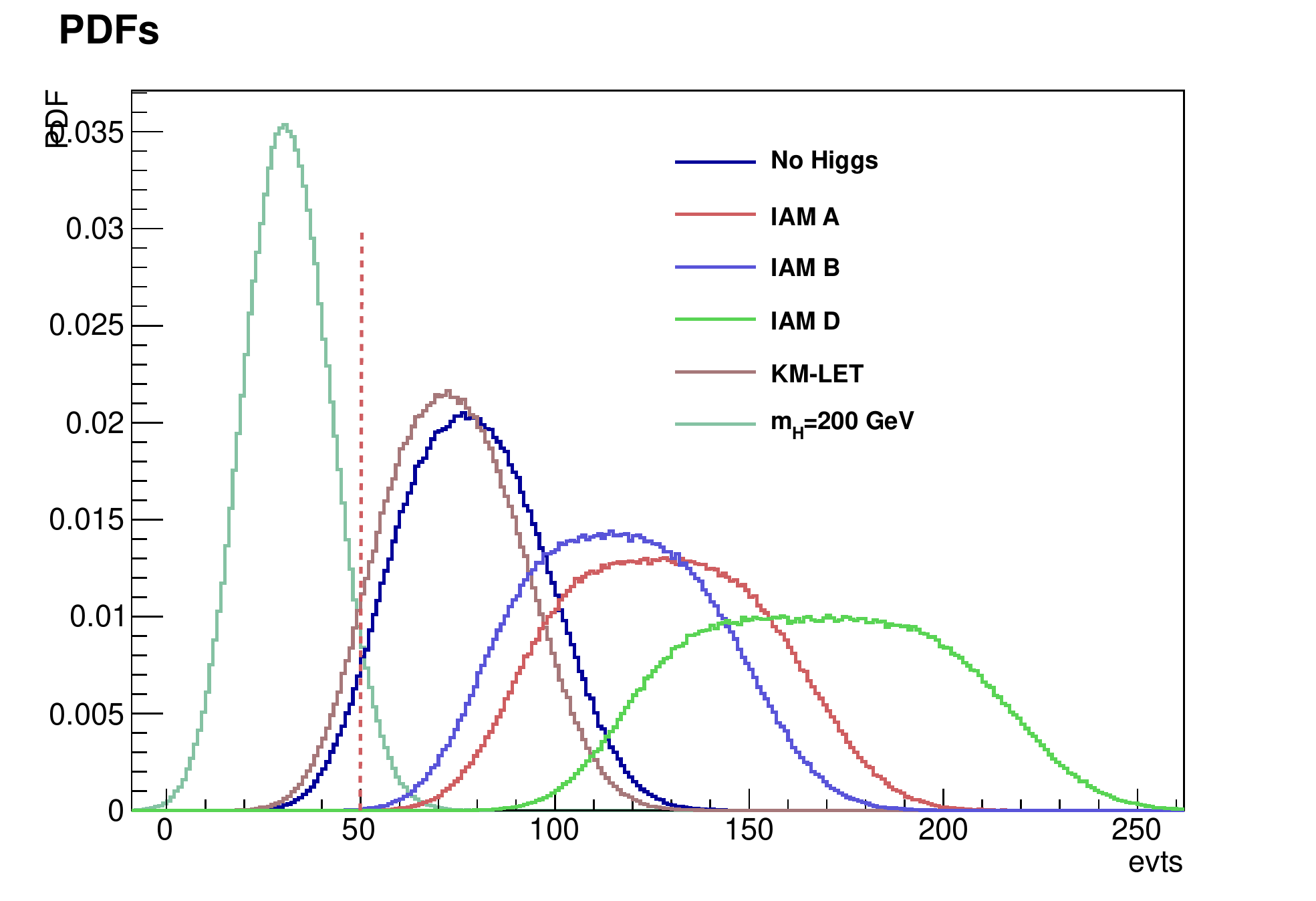}
\hspace*{-0.02\textwidth}
\includegraphics*[width=0.52\textwidth,height=6cm]{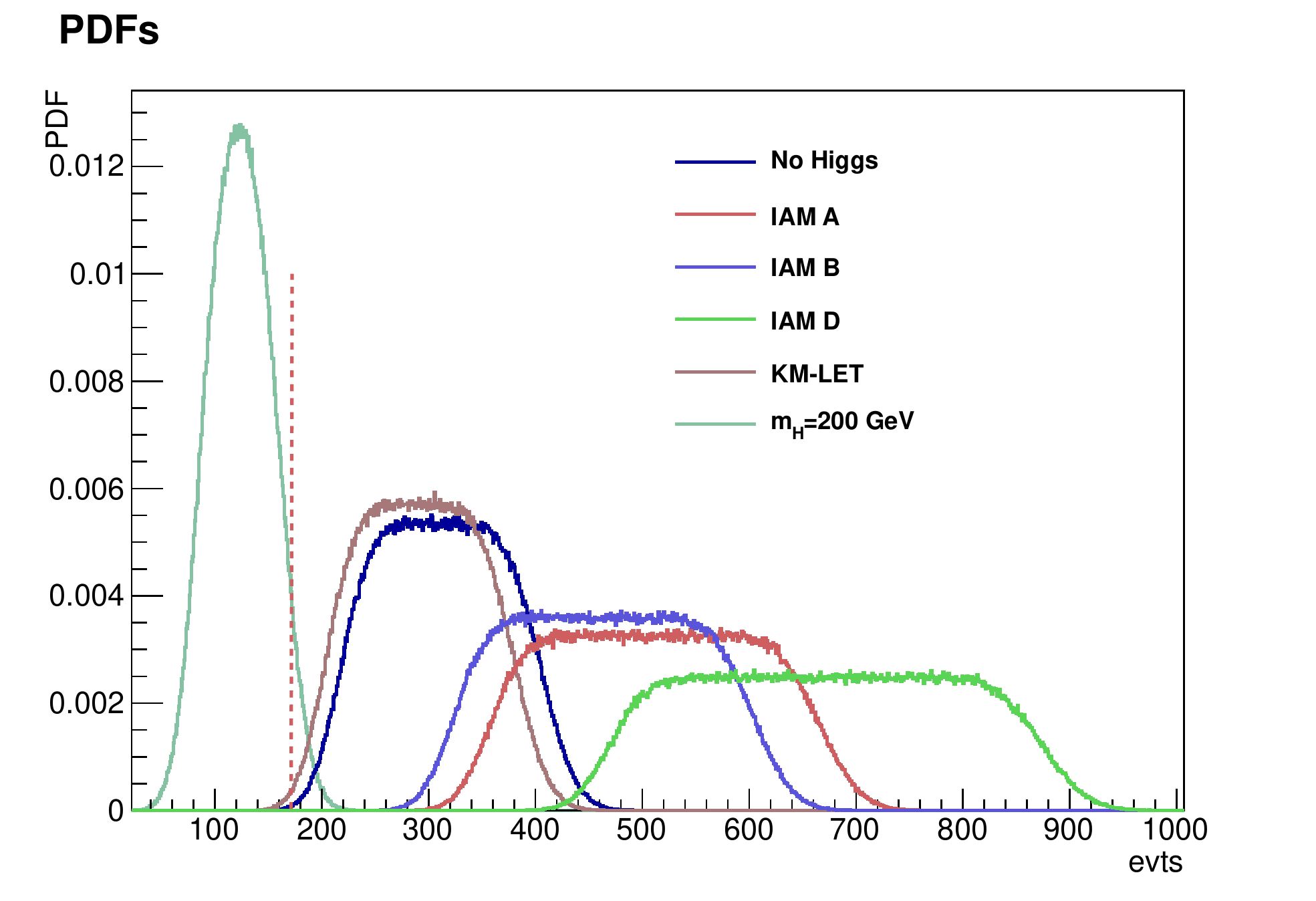}
}
\caption{PDFs of the number of events in the $\ell^\pm\nu+4j$ channel
for $M_{cut}=800$ GeV, $L=50\ifb$ (left) and 
$M_{cut}=800$ GeV, $L=200\ifb$ (right).}
\label{fig:PDF4jlv}
\end{figure}

Examining the PDF distributions of \fig{fig:PDF4jlv}, one realizes that
the situation is similar to the one already discussed for
the $2jW^+W^-\ra 2j\ell^+\ell^-\nu\bar{\nu}$
channel: the models with the lightest resonances are the ones which can be
most easily distinguished from the Standard Model with a light Higgs. The KM-LET
has the smallest cross section of all unitarized models.
It is however well separated by the SM case especially with a luminosity of $200\ifb$.
The no-Higgs model behaves similarly to the KM-LET one, confirming that,
even if unphysical, this model has good properties as a benchmark case.

\begin{table}[ht!]
\centering
\begin{tabular}{|c|c|c|c|c|c|c|}
\hline
$L (\ifb)$ &  $M_{cut}$ (GeV) & no--Higgs& KM--LET & IAM A  & IAM B    & IAM D \\
\hline	
50  & 800   & 94.51\% & 91.03\% & 99.99\% & 99.97\% & 100\%\\
200  & 800   & 99.93\% & 99.64\% &100\% &100\% & 100\%\\
\hline
\end{tabular}
\caption{PBSM@95\%CL, defined in \subsect{subsec:2jWW}, for the $\ell^\pm\nu+4j$ channel.}
\label{tab:pbsm_4jlv}
\end{table}

Finally, the number of events is rather large already at  $L=50\ifb$ and the
discriminating power described by the PBSM@95\%CL values of \tbn{tab:pbsm_4jlv}
is rather encouraging.

\subsubsection{The $\ell^+\ell^-+4j$ channel}

This channel is in many respects similar to the one we have just considered.
The final state in this case corresponds to ZW or ZZ production in the
boson boson scattering processes. There are no contributions from
$t\bar{t}+2j$ but the $Z+4j$ background is large. It has been evaluated with
the running scale $Q=\hat{H}_T'/2$ as before.

In \fig{fig:4jll_MWV} we show the distribution of the
invariant mass of the two final state bosons for the different models.
After all cuts, the signal cross sections in this channel are about five times
smaller than those of the $4j\ell\nu$ channel, as shown in \tbn{tab:res_4jll}.
This implies that the 
discriminating power for new physics models is sensibly lower, as can be
seen from \fig{fig:PDF4jll} and \tbn{tab:pbsm_4jll}.

\begin{figure}[h!]
\centering
\includegraphics*[width=0.62\textwidth,height=6cm]{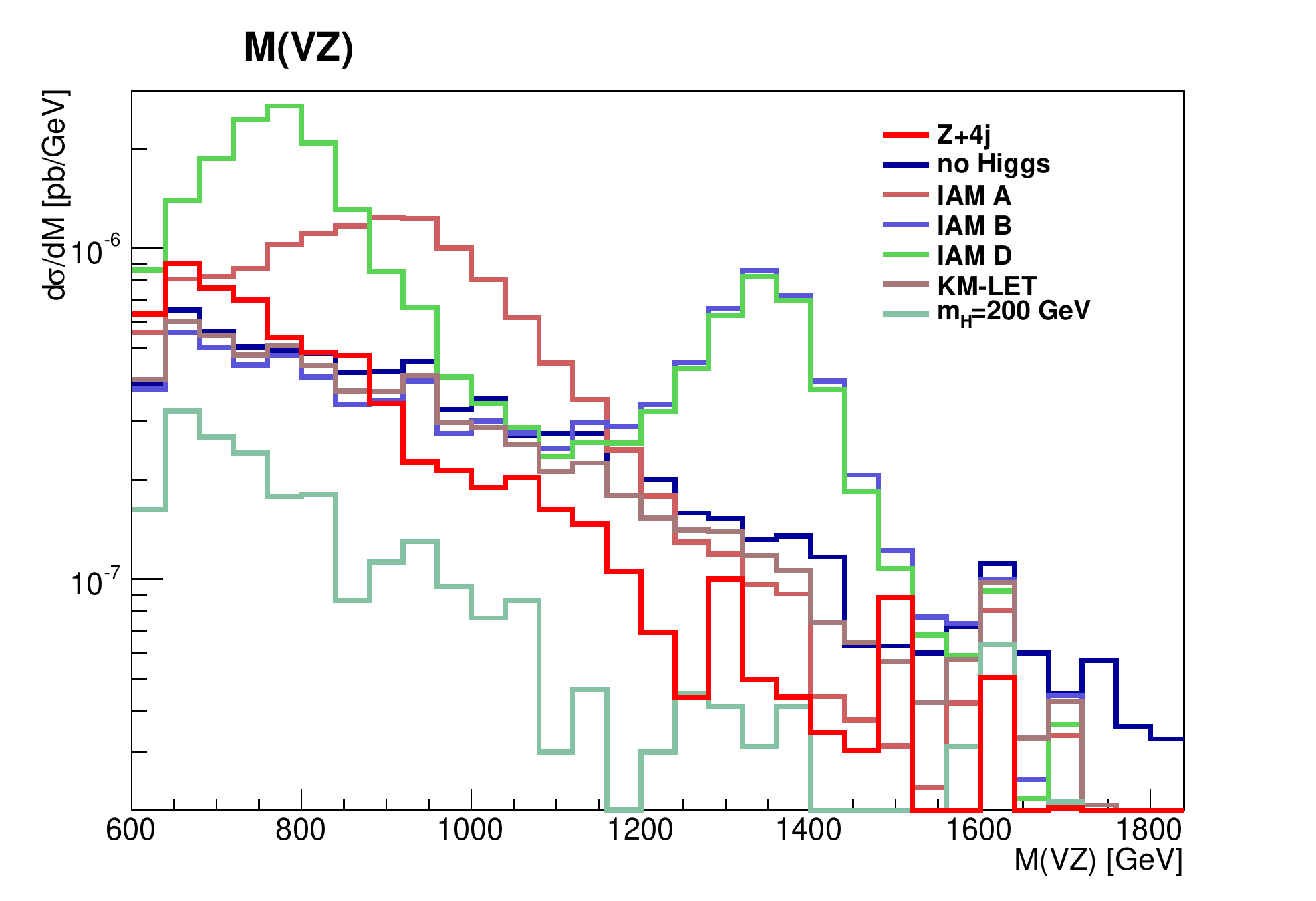}
\caption{Distributions of the mass of the $VZ$-system for the $\ell^+\ell^-+4j$ channel  
with full set of cuts, \tbn{tab:cuts_basic} and \tbn{tab:addcuts}. 
}
\label{fig:4jll_MWV}
\end{figure}

\begin{table}[ht!]
\centering
\begin{tabular}{|c|c|c|c|c|c|c|c|}
\hline
$M_{cut}$      &no--Higgs& KM--LET 	& IAM A	& IAM B	& IAM D & SM    & $Z+4j$ \\ 
\hline
600		& .308  & .274  & .532  & .388  & .795  & .0969 & .269 \\
800             & .204  & .173  & .369  & .294  & .424  & .0500 & .128  \\
1000            & .120  & .0975 & .138  & .223  & .211  & .0258 & .059  \\
1200            & .0657 & .0511 & .0396 & .167  & .156  & .0154 & .027 \\
\hline
\end{tabular}
\caption{Total cross section (in fb) for the $\ell^+\ell^-+4j$ channel 
with the full set of cuts in \tbn{tab:cuts_basic} and \tbn{tab:addcuts}
in function of the minimum $j_cj_c\ell\ell$ invariant mass (in GeV).
}
\label{tab:res_4jll}
\end{table}

\begin{figure}[t!]
\centering
\subfigure{
\includegraphics*[width=0.52\textwidth,height=6cm]{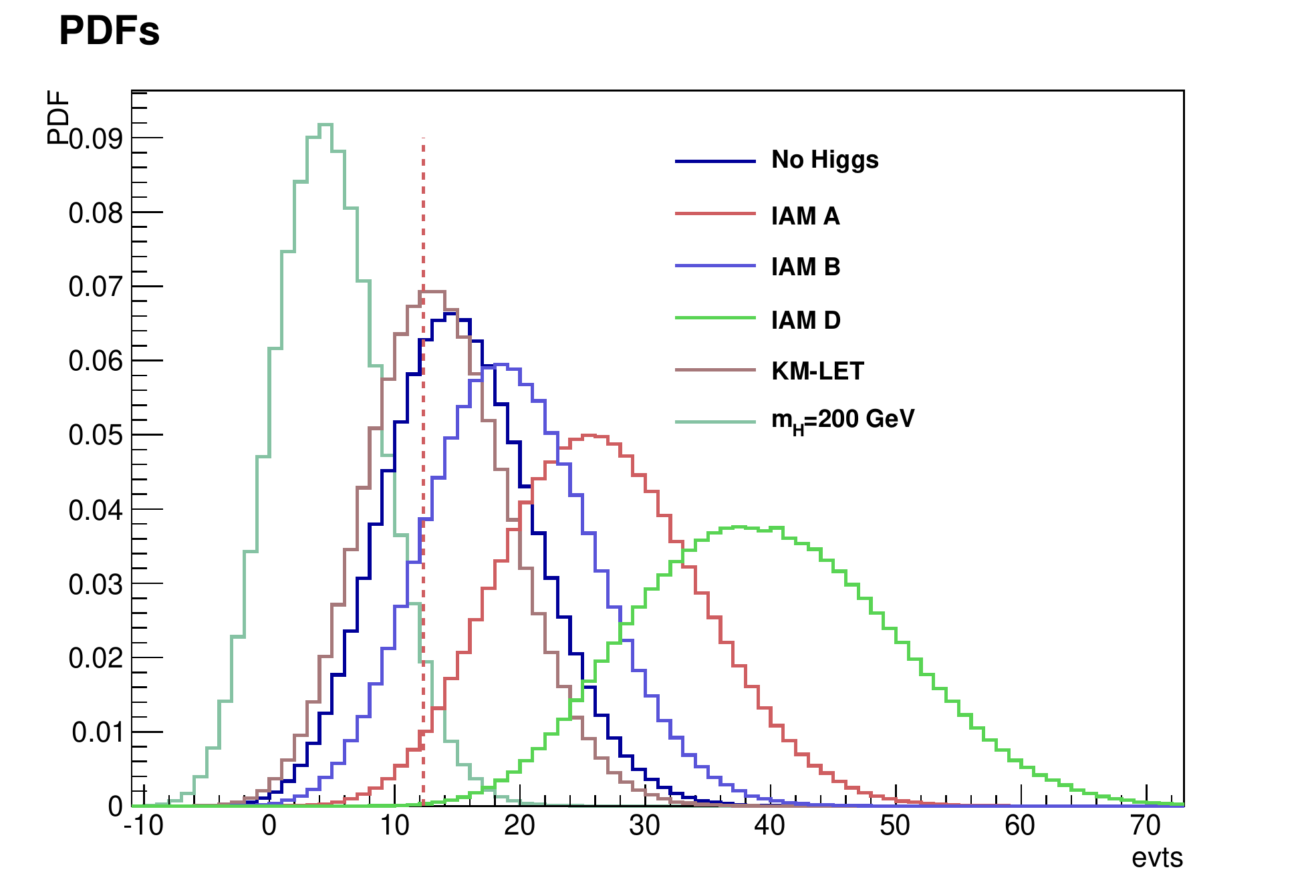}
\hspace*{-0.02\textwidth}
\includegraphics*[width=0.52\textwidth,height=6cm]{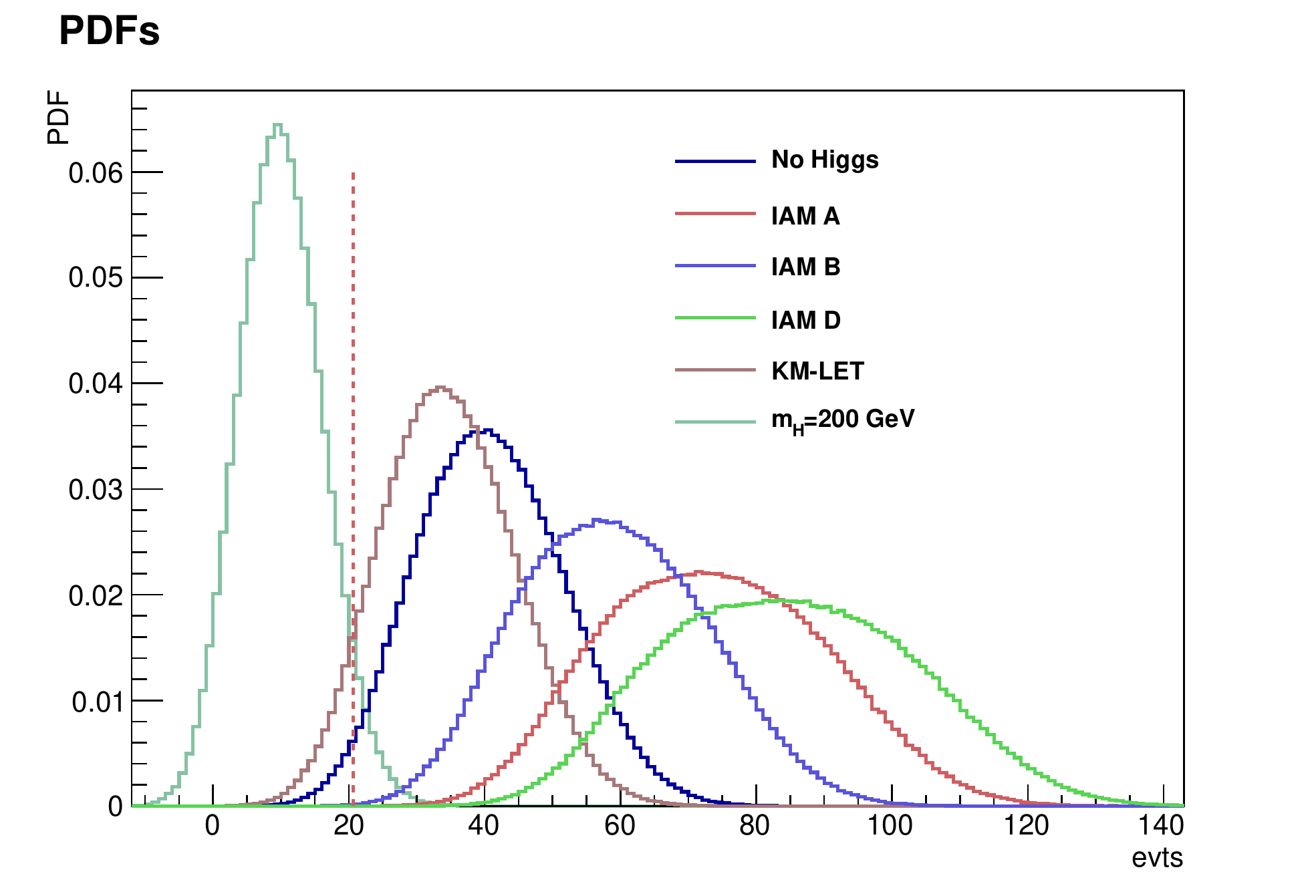}
}
\caption{PDFs of the number of events in  the $\ell^+\ell^-+4j$ channel
for $M_{cut}=600$ GeV, $L=50\ifb$ (left) and $M_{cut}=800$ GeV, 
$L=200\ifb$ (right).  }
\label{fig:PDF4jll}
\end{figure}

\begin{table}[ht!]
\centering
\begin{tabular}{|c|c|c|c|c|c|c|}
\hline
$L (\ifb)$ &  $M_{cut}$ (GeV)	& no--Higgs& KM--LET & IAM A  & IAM B    & IAM D \\
\hline	
50      &    600	& 68.36\% & 57.90\% & 97.43\% & 85.61\% & 99.95\%\\
200     &    800 & 97.70\% &92.81\% & 99.99\% &99.97 .\% & 100\%\\
\hline
\end{tabular}
\caption{PBSM@95\%CL, defined in \subsect{subsec:2jWW}, for the $\ell^+\ell^-+4j$ channel
for $L=50\ifb$ and $L=200\ifb$.}
\label{tab:pbsm_4jll}
\end{table}

The pattern of the distributions in \fig{fig:PDF4jll} is analogous
to the one in the $4j\ell\nu$ channel: models with light resonances are easily
distinguished from the SM and models without resonant states give smaller predictions
which are in rough agreement among themselves.
With $L=200\ifb$ all models have a high probability of producing a number of
events larger than the 95\%CL for the SM. 
The KM-LET model, which is the one least likely to be detected sports
a PBSM@95\%CL of 93\%.
At $L=50\ifb$ the PBSM@95\%CL for the KM-LET case drops to about 60\%.

\subsubsection{The $3\ell\nu+2j$ channel}

This final state does not suffer from the huge backgrounds stemming from $t\overline{t}$ or 
$V+4j$ production.
The final state corresponds to ZW production in the underlying
scattering process. Hence scalar resonances cannot be formed.
This is clearly visible in \fig{fig:2j3l_MWV} where we show on the left 
the boson boson invariant mass distribution and on the right the PDF for 
$L=200\ifb$ and $M_{cut}=600$ GeV. One notices immediately, comparing with the
results obtained for the two previous
channels, that the PDF for the IAM A model, which develops a scalar resonance, is now
closer to the SM predictions than the KM-LET model.

\begin{figure}[h!]
\centering
\includegraphics*[width=0.5\textwidth,height=6cm]{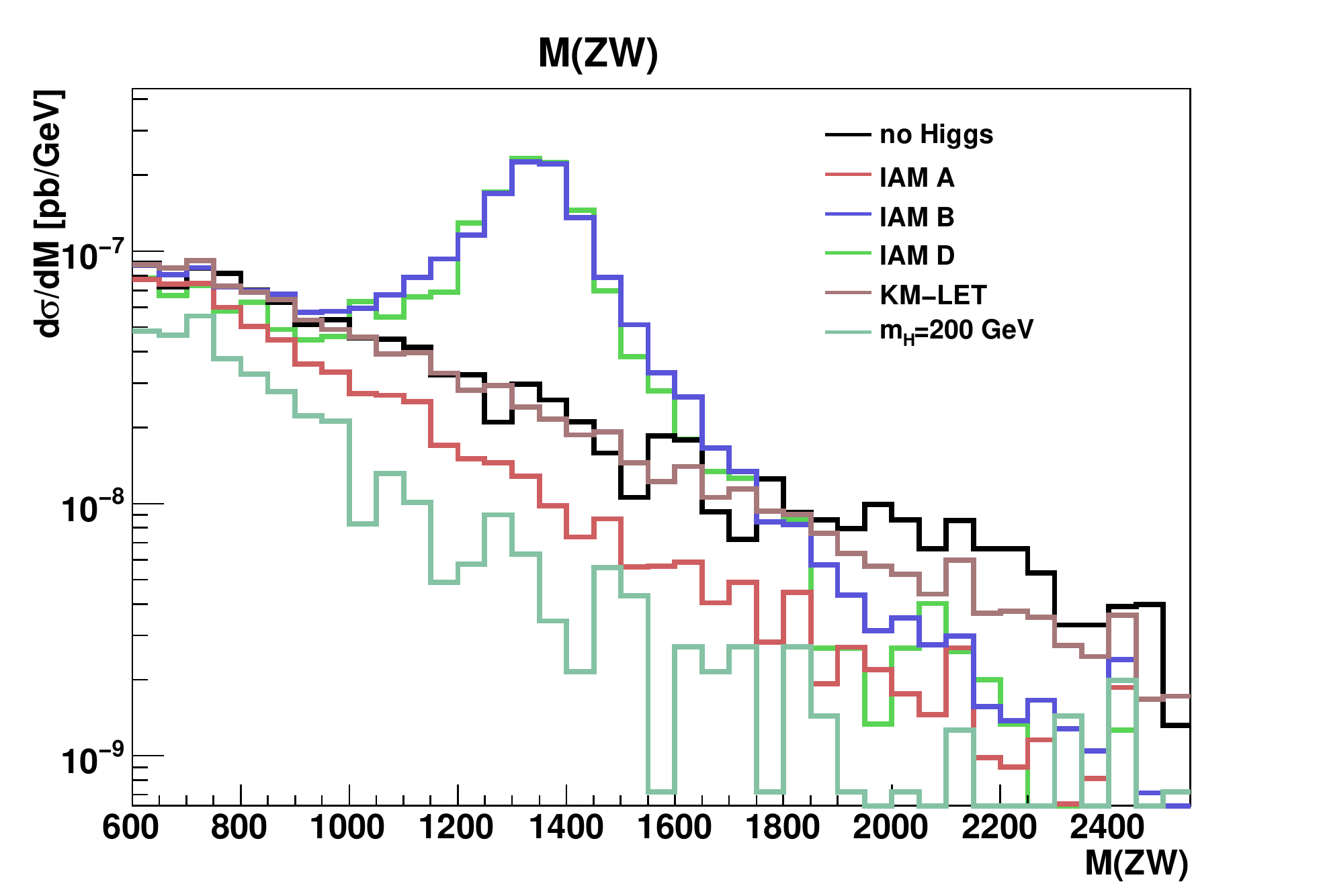}
\hspace*{-0.02\textwidth}
\includegraphics*[width=0.5\textwidth,height=6cm]{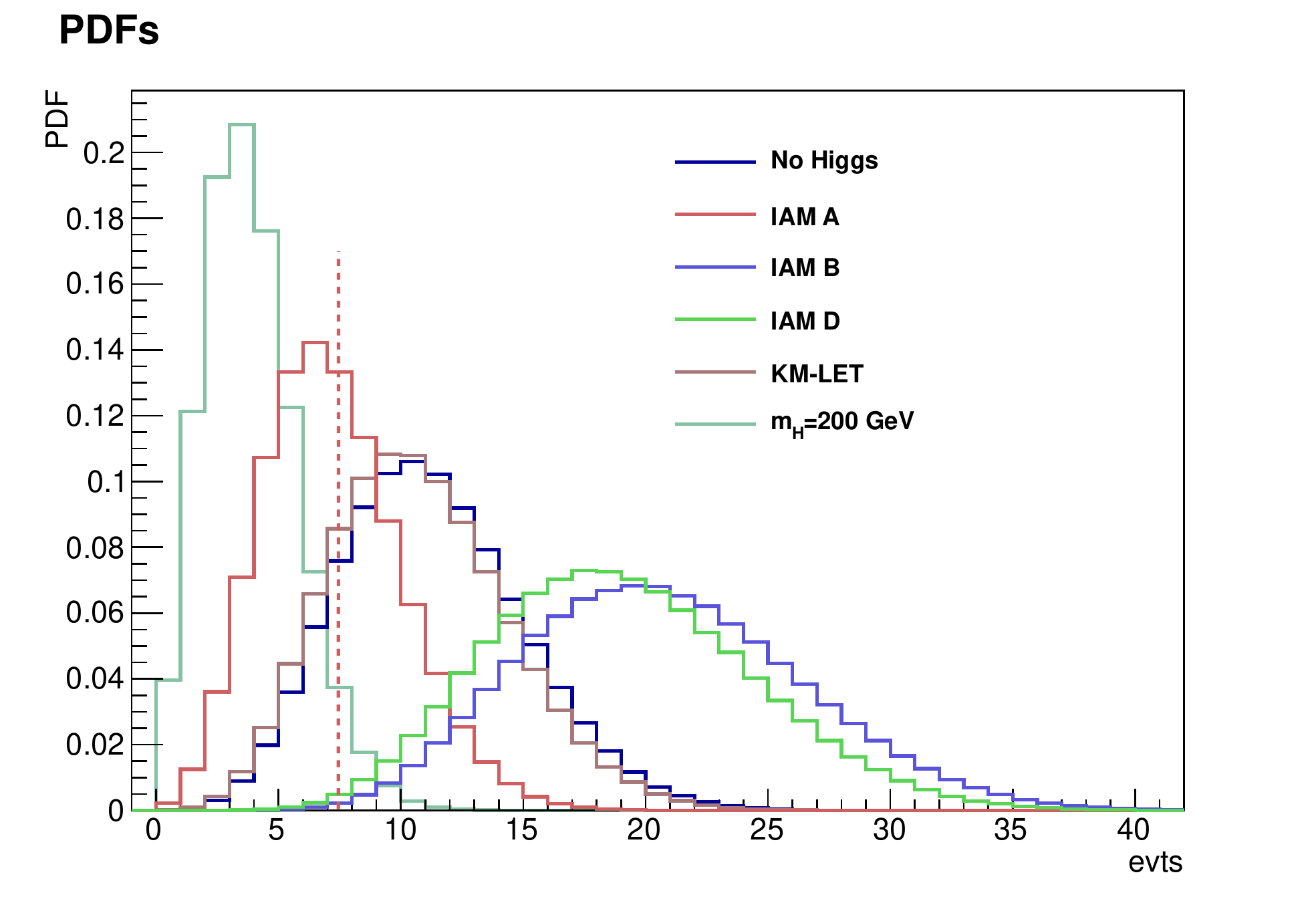}
\caption{Distributions of the mass of the $WZ$-system for the  $3\ell\nu+2j$ channel
with full set of cuts, \tbn{tab:cuts_basic} and \tbn{tab:addcuts} (left).
PDFs of the number of events for $M_{cut}=600$ GeV and $L=200\ifb$ (right).}
\label{fig:2j3l_MWV}
\end{figure}

The total cross sections as a function of the boson boson invariant mass
lower cut are given in \tbn{tab:res_3lv}. The reported values are now in 
attobarns and this is reflected in a small expected number of events at the LHC.
The  PBSM@95\%CL are shown in 
\tbn{tab:pbsm_2j3l}.  With $50\ifb$ of data only the IAM B and IAM D models have a
reasonable chance of manifesting themselves in sizable deviations from the SM.
At $L=200\ifb$ these two models would be almost certainly revealed while
the PBSM@95\%CL for  both the no-Higgs and the KM-LET models are above 80\%.

\begin{table}[ht!]
\centering
\begin{tabular}{|c|c|c|c|c|c|c|}
\hline
$M_{cut}$     &no--Higgs& KM-LET   & IAM A& IAM B & IAM D & SM    \\ 
\hline
600		& 53.8 & 51.2 & 33.5 & 101.  & 92.6  & 17.0 \\
800             & 37.6 & 34.3 & 19.2 & 84.6  & 78.8  & 8.92 \\
1000            & 25.6 & 22.5 & 11.0 & 72.0  & 68.7  & 4.42 \\
1200            & 17.3 & 14.7 & 6.21 & 57.1  & 56.1  & 2.38 \\
\hline
\end{tabular}
\caption{Total cross section (in $ab$) for the $3\ell\nu+2j$ channel 
with the full set of cuts in \tbn{tab:cuts_basic} and \tbn{tab:addcuts}
in function of the minimum $3\ell\nu$ invariant mass (in GeV).
}
\label{tab:res_3lv}
\end{table}

\begin{table}[ht!]
\centering
\begin{tabular}{|c|c|c|c|c|c|c|}
\hline
$L (\ifb)$ &  $M_{cut}$ (GeV) & no--Higgs& KM--LET & IAM A  & IAM B    & IAM D \\
\hline	
50      & 600	& 46.30\% & 43.27\% & 21.51\% & 83.71\% & 79.51\%\\
200     & 600	& 84.12\% & 80.81\% & 43.47\% & 99.73\% & 99.38\%\\
\hline
\end{tabular}
\caption{PBSM@95\%CL, defined in \subsect{subsec:2jWW}, for the  $3\ell\nu+2j$ channel.}
\label{tab:pbsm_2j3l}
\end{table}

\subsubsection{The $4\ell+2j$ channel}

The present channel has some similarities with the $3\ell\nu+2j$ channel
above. In some sense it is complementary to it. 
In fact it corresponds to the formation of a ZZ pair
and this implies that no vector resonance can be formed in this channel.
In \fig{fig:2j4l_MZZ} the ZZ invariant mass distribution is shown on the left
while the PDF for $L=200\ifb$ and $M_{cut}=500$ GeV are on the right.
\fig{fig:2j4l_MZZ} clearly shows that the models which produce scalar resonances are
well separated from the SM. On the contrary, the IAM B model, in which only a vector
resonance is present,
appears  even more difficult to disentangle from the light-Higgs scenario
than the non resonant ones.

\begin{figure}[h!]
\centering
\includegraphics*[width=0.5\textwidth,height=6cm]{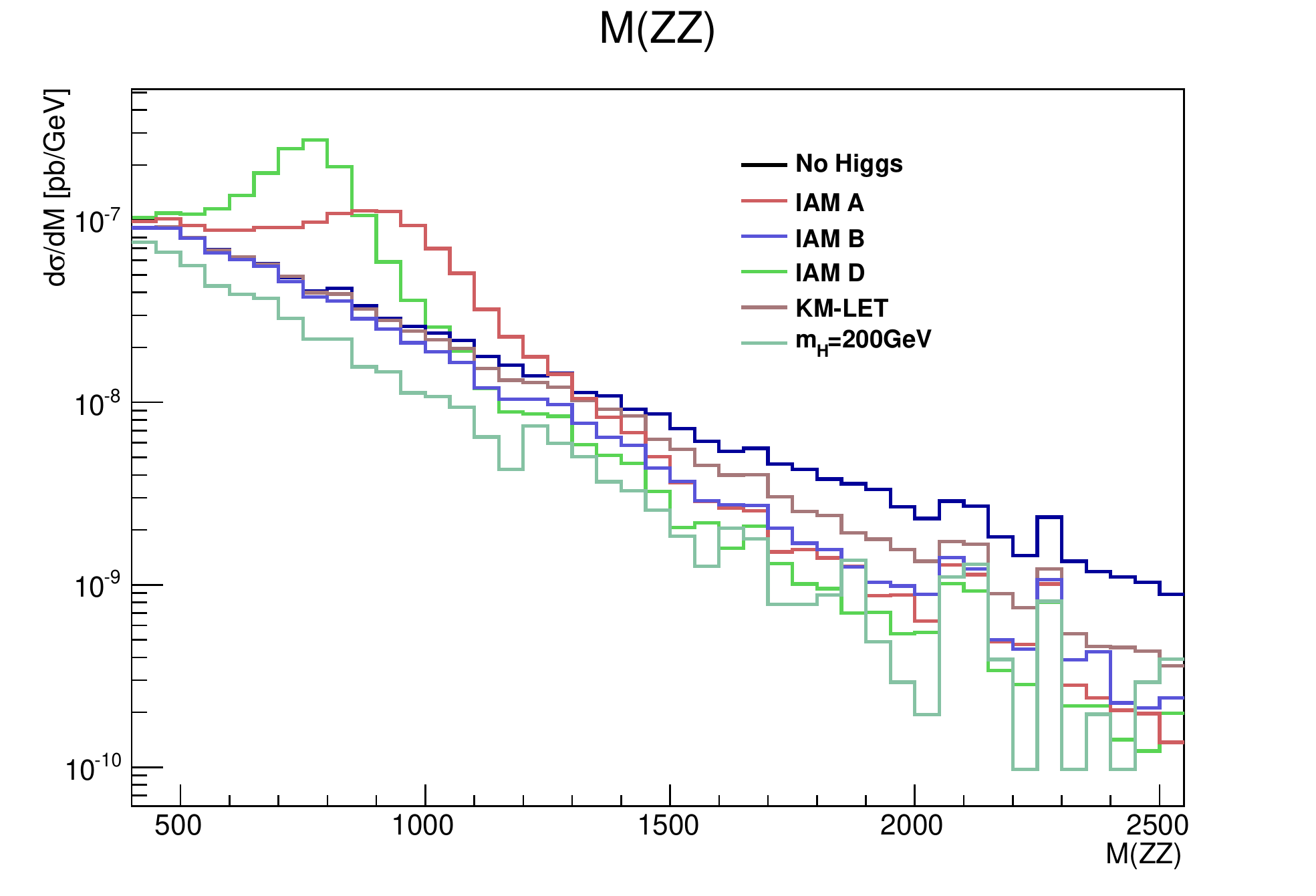}
\hspace*{-0.02\textwidth}
\includegraphics*[width=0.5\textwidth,height=6cm]{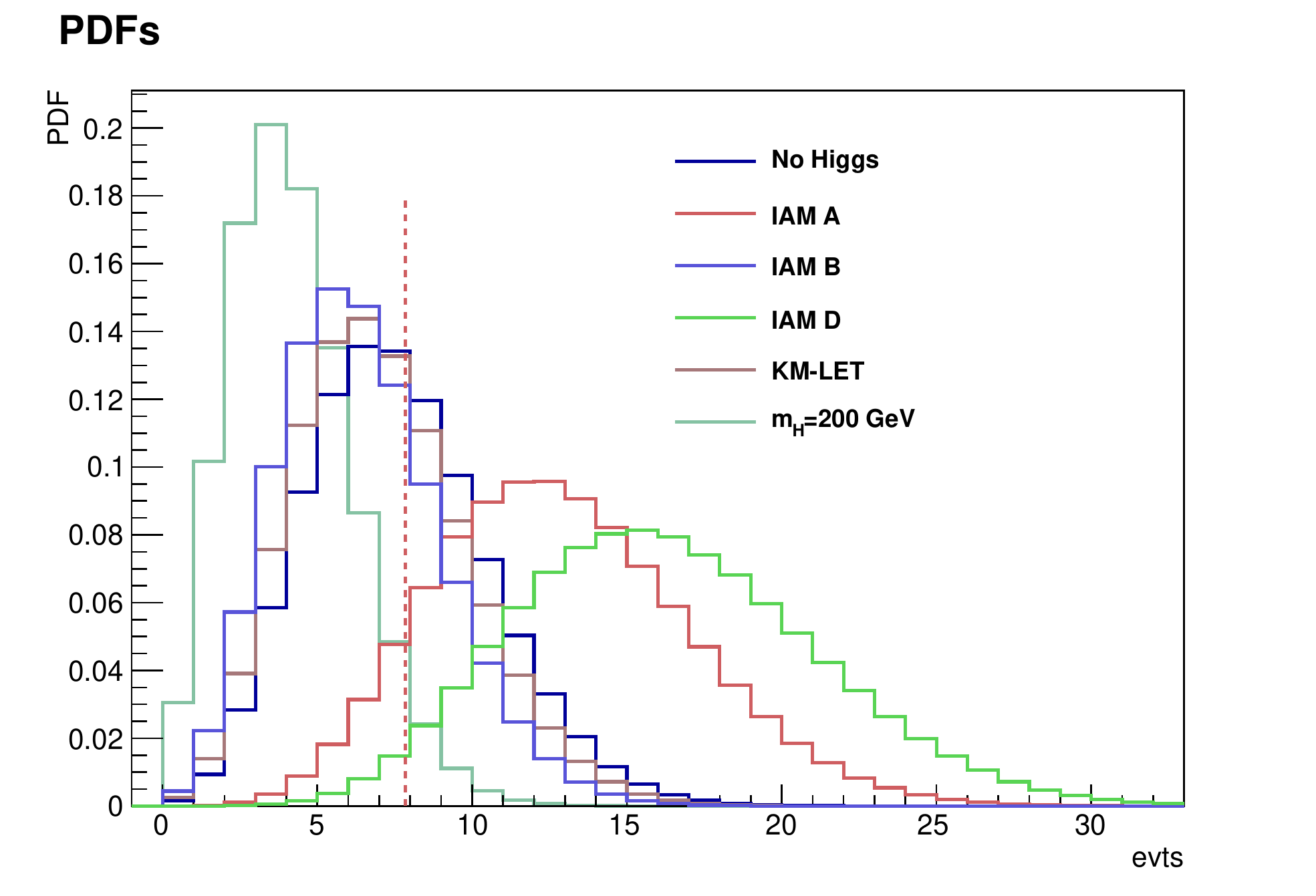}
\caption {Distributions of the mass of the $ZZ$-system for the $4\ell+2j$ channel
with full set of cuts, \tbn{tab:cuts_basic} and \tbn{tab:addcuts} (left).
PDFs for $M_{cut}=500$ GeV and $L=200\ifb$ (right).
}
\label{fig:2j4l_MZZ}
\end{figure}

\begin{table}[ht!]
\centering
\begin{tabular}{|c|c|c|c|c|c|c|}
\hline
$M_{cut}$     &no--Higgs& KM--LET   & IAM A& IAM B & IAM D & SM    \\ 
\hline
400		& 44.7 & 41.9 & 72.0 & 38.5  & 89.6  & 25.5 \\
500             & 35.6 & 32.8 & 62.0 & 29.5  & 79.0  & 18.4 \\
600             & 28.2 & 25.4 & 52.9 & 22.2  & 67.8  & 13.5 \\
700             & 22.2 & 19.4 & 44.0 & 16.4  & 52.0  & 9.64 \\
800		& 17.8 & 14.9 &	34.6 & 12.2  & 26.0  & 7.09  \\
900		& 14.0 & 11.4 &	23.6 & 8.97  & 10.9  & 5.19  \\
\hline
\end{tabular}
\caption{Total cross section in $ab$ for the $4\ell+2j$ channel 
with the full set of cuts in \tbn{tab:cuts_basic} and \tbn{tab:addcuts}
in function of the minimum $4\ell$ invariant mass (in GeV). 
}
\label{tab:res_4l}
\end{table}

In \tbn {tab:res_4l} the values of the cross sections in attobarns are reported.
These very modest cross sections imply that no stringent cut can be imposed
and as a consequence the discriminatory power of this channel is smaller than that of
all other final states presented in this paper, as can be seen from \tbn{tab:pbsm_4l}.
Only at $L=200\ifb$ this channel could be useful for the IAM A and IAM D models.

\begin{table}[ht!]
\centering
\begin{tabular}{|c|c|c|c|c|c|c|}
\hline
$L (\ifb)$ &  $M_{cut}$ (GeV) & no--Higgs& KM--LET & IAM A  & IAM B    & IAM D \\
\hline	
200     & 500		& 43.86\% & 36.31\% & 89.60\% & 27.41\% & 97.36\% \\
\hline
\end{tabular}
\caption{PBSM@95\%CL, defined in \subsect{subsec:2jWW}, for the $4\ell+2j$ channel.}
\label{tab:pbsm_4l}
\end{table}


\section{Conclusions}

\label{sec:conclusions}

We have described in some detail how  Unitarized Models have been 
implemented in the \Phantom MonteCarlo. We have then compared
the predictions of this new 
implementation which makes use of  full six fermion final state matrix 
elements calculations with those obtained with old EVBA methods
and presented in the literature.  The agreement between the two calculations
depends rather strongly on the separation between the two tag jet, a variable which
cannot be controlled in EVBA. The shape of the distribution of the invariant
mass of the two vector boson system is in any case quite different, in particular for
smaller invariant masses.

Using the new tool we have adressed the important issue
of the  possibility of uncovering new physics signals in vector boson scattering
processes at LHC. For this we have considered some typical examples of unitarized
models. These can be divided in two sets: those which predict  the formation of resonances,
which in the cases we have examined can be scalar or vector in nature,
and those in which no resonant state is formed.

Our analysis has concentrated on counting experiments which look for an excess
of boson boson scattering events compared with SM expectations in the large invariant
mass region.
Our results show that these experiments can detect new physics at 14 TeV for all considered scenarios
using a luminosity of 200 $\ifb$. At 50 $\ifb$ instead the results
depend on the particular channel and scenario under consideration. 
Since models with no resonance, models with scalar resonances and models
with vector resonances behave differently in different final states
we stress the importance of carefully analyzing
all possible channels and not only the ones with largest cross section or smallest background.


\section *{Acknowledgments}

A.B. wishes to thank the Dep. of Theoretical Physics of Torino University
for support. \\
This work has been supported by MIUR under contract 2008H8F9RA\_002 . 
\newpage

\appendix

\section{Selection cuts}

\label{AppendixA}

\begin{table}[th!]
\begin{center}
\begin{tabular}{|l|c|c|c|c|c|c|c|}
\hline
\hfill  Processes  &$2j\ell^+\ell '^-\nu\bar\nu$& $2j\ell^+\ell^-\nu\bar\nu$ 
                                 &$2j\ell^\pm\nu \ell^\pm\nu$ 
						&4j$\ell\nu$&$4j\ell\ell$&$2j3\ell\nu$&$2j4\ell\nu$\\
     Cuts                   &$(W^+W^-)$ &$(ZZ)$   & & & & & \\
\hline
$|\eta(\ell^\pm)|<$ 	& 2.0	& 	& 	& 2.0 	&	& 2.0	& 	\\
\hline
$M(j_fj_b)>$		& 1000	& 800	&	& 1000 	& 1000	& 1000	& 800	\\
\hline
$|\Delta \eta(j_fj_b)|>$& 4.8	& 4.5	& 4.5	& 4.8 	& 4.8	& 4.8	& 	\\
\hline
$p_T(j_c) >$		& 	& 	& 	& 70 	& 60	& 	&	\\
\hline
$p_T(j_c j_c) > $	& 	&	&	& 	& 200	& 	&\\
\hline
$p_T(\ell\nu) >$	& 	&	& 	& 200 	&	& 200	&	\\
\hline
$p_{Tmiss} >$		& 	& 120	& 	& 100 	&	&	&	\\
\hline
$p_T(\ell^+\ell^-)>$	& 	& 120 	& 	& 	& 200	& 200	& 100	\\			
\hline
$p_T(\ell) > $		& 	& 	& 50 	&	&	&	&	\\
\hline
2$min{p_T(j)} <$		&	&	& 120 	&	&	&	&	\\
\hline
$E(j)>$			& 180 	& 	& 	&	&	&	&	\\
\hline
$max|\eta(j)|>$		& 2.5 	&	& 2.5	& 	& 2.8	&	&	\\
\hline
$|\eta(j)|>$		& 1.3 	& 1.9	&  	&	&	& 1.2	&	\\
\hline
$|\Delta\eta(Vj)|>$	& 	&	& 	& 0.6 	& 1.1	& 1.5	&	\\
\hline
$\Delta\eta(\ell j)>$	& 0.8 	& 1.3	& 	&	&	&	&	\\
\hline
$\Delta R (\ell j)>$	& 1 	& 	& 1.5	&	&	&	&	\\
\hline
$\Delta R (Z j)>$	&	&	&	&	&	& 	& 1 \\
\hline
$M(\ell j)>$		& 180 	&	& 	&	&	&	&	\\
\hline
$M(Vj)>$		&	&	&	&	& 300 	& 	&\\
\hline
$|\vec{p}_T(\ell_1)-\vec{p}_T(\ell_2)|>$
			& 220 	& 	& 150	&	&	&	&	\\
\hline
$|\vec{p}_T(\ell^+\ell^-)-\vec{p}_T^{\,miss}|>$ 
			& 	& 290 	& 	&	&	&	&	\\
\hline
$\cos (\delta\phi_{\ell\ell})<$
			& -0.6 	& 	& -0.6	&	&	&	&	\\
\hline
$\cos (\delta\phi_{ZZ})<$
			&  	& 	& 	&	&	&	& -0.4	\\
\hline
$\Delta R(\ell^+\ell^-)<$& 	&	&	& 	& 1.0	& 	&\\
\hline
\end{tabular}
\caption{Additional selection cuts for the various channels.
All masses, momenta and energies are expressed in GeV.}
\label{tab:addcuts}\end{center}
\end{table}

\clearpage


\bibliographystyle{JHEP}
\bibliography{mybib}

\end{document}